\documentclass[aps,showpacs,preprintnumbers,amsmath,amssymb]{revtex4}
\usepackage{dcolumn}
\usepackage[dvips]{epsfig}
\usepackage{graphicx}
\usepackage{amssymb}
\usepackage{color}
\usepackage{enumerate}

\begin{document}
\baselineskip=0.8 cm
\title{\bf Rotating charged black hole with  Weyl corrections}

\author{Songbai Chen\footnote{csb3752@hunnu.edu.cn}, Jiliang Jing
\footnote{jljing@hunnu.edu.cn}}
%\email{csb3752@163.com}

\affiliation{Institute of Physics and Department of Physics, Hunan
Normal University,  Changsha, Hunan 410081, People's Republic of
China \\ Key Laboratory of Low Dimensional Quantum Structures \\
and Quantum Control of Ministry of Education, Hunan Normal
University, Changsha, Hunan 410081, People's Republic of China}

\begin{abstract}
\baselineskip=0.6 cm
\begin{center}
{\bf Abstract}
\end{center}
We present firstly a four-dimensional spherical symmetric black hole with small Weyl corrections and find that with increasing Weyl corrections the region of the event horizon existence for the black hole in the parameter space increases for the negative Weyl coupling parameter and decreases for the positive one. Moreover, we also obtain a rotating charged black hole with weak Weyl corrections by the method of complex coordinate transformation. Our results show that the sign of Weyl coupling parameter $\alpha$ yields the different spatial topology of the event horizons for the black hole with its parameters lied in some special regions in the parameter space.
We also analyze the dependence of the ergosphere on the Weyl coupling parameter $\alpha$ and find that with the increase of the Weyl corrections the ergosphere in the equatorial plane becomes thick for a black hole with $\alpha>0$, but becomes thin in the case with $\alpha<0$, which means that
the energy extraction become easier in the background of a black hole with the positive Weyl coupling parameter, but more difficult in the background of a black hole with the negative one.

\end{abstract}

 \maketitle
\newpage
\section{Introduction}

The generalized Einstein-Maxwell theories have received a lot of attention
recently because it contains higher
derivative interactions and carries more information about the electromagnetic field. The study of such kind of generalized Einstein-Maxwell theories could help us to explore the full properties and effects of the electromagnetic fields. In general, the generalized Einstein-Maxwell theories can be classified into two classes. The first class is minimally coupled gravitational-electromagnetism in which there is no coupling
in the Lagrangian between the Maxwell part and the curvature part.
One of the interesting generalized Einstein-Maxwell theory belong to this class is Born-Infeld theory \cite{Born} which removes the divergence of the electron's self-energy in the classical electrodynamics and possesses good physical properties including the absence of shock waves and birefringence phenomena
\cite{Boillat}. Moreover, it is found that Born-Infeld theory  enjoys an electric-magnetic duality \cite{Gibbons} and can describe gauge fields on a D-brane which arises from attached open strings \cite{Fradkin}.
The second class includes the nonminimal coupling between the gravitational and electromagnetic fields in the Lagrangian \cite{Balakin,Faraoni,Hehl}.  This class is of great interest because
the appearance of the nonminimal couplings in the Lagrangian modifies the coefficients of the second-order derivatives both in the Maxwell and Einstein equations, which could affect the propagation of gravitational and electromagnetic waves in the spacetime and may yield time delays in the arrival of those waves \cite{Balakin}. In the evolution of the early Universe, such a kind of coupled terms may result in electromagnetic quantum fluctuations and lead to the inflation \cite{Turner,Mazzitelli,Lambiase,Raya,Campanelli}. Recent investigations also show that these cross-terms have been used as attempts to explain the large scale magnetic fields observed in clusters of galaxies \cite{Bamba,Kim,Clarke}.

It is of interest to search the solutions of black holes  in the generalized Einstein-Maxwell theories and to probe how the generalized electrodynamics modify the properties of the black hole. In the frame of the Born-Infeld theory, the electrically charged black hole solutions were obtained in \cite{BIBH1,BIBH2,BIBH3}, which displays that the black hole singularity in this theory is weaked from that of usual Reissner-Nordstr\"{o} m black hole.
In order to avoid the black hole singularity problem, some regular models of black holes have been proposed in \cite{Bardeen,Borde}, which are called as Bardeen black holes. The Bardeen black holes can be interpreted as the solution to a nonlinear magnetic monopole with a mass $M$ and a charge $q$ \cite{Beato1}. The Bardeen black holes are also generalized to the model with four specific parameters \cite{Beato}. Recently, a large class of black hole solutions have been constructed in the power Maxwell theory \cite{pwbh1,pwbh2,pwbh3,pwbh4} in which the Maxwell action takes as power-law function of the form $\mathcal{L}=-\beta (F_{\mu\nu}F^{\mu\nu})^{k}$, where $\beta$ is a coupling constant and $k$ is a power parameter. It is found that the asymptotic behavior of the solution depends heavily on the value of the power parameter $k$.
Moreover, the black hole solution have been considered in the modified Maxwell field including the nonminimal coupling between the gravitational and electromagnetic fields \cite{Balakin1}. It is shown that these coupled terms modify the electromagnetic and gravitational structure of a charged black hole.

One of simple generalized electromagnetic theories is the electrodynamics with Weyl corrections which involves a coupling between
the Maxwell field and the Weyl tensor \cite{Weyl1,Drummond}.
In this theory, the Lagrangian density of the electromagnetic
field is modified as
\begin{eqnarray}\label{LEM}
L_{EM}=-\frac{1}{4}\bigg(F_{\mu\nu}F^{\mu\nu}-4\alpha
C^{\mu\nu\rho\sigma}F_{\mu\nu}F_{\rho\sigma}\bigg),
\end{eqnarray}
where $F_{\mu\nu}$ is
the usual electromagnetic tensor, which is related to the electromagnetic
vector potential $A_{\mu}$ by $F_{\mu\nu}=A_{\nu;\mu}-A_{\mu;\nu}$. The coefficient $\alpha$ is a coupling constant with dimensions of length squared and the
tensor $C_{\mu\nu\rho\sigma}$ is so-called Weyl tensor, which is related to the Riemann tensor $R_{\mu\nu\rho\sigma}$, the Ricci tensor $R_{\mu\nu}$ and the Ricci scalar $R$ by
\begin{eqnarray}
C_{\mu\nu\rho\sigma}=R_{\mu\nu\rho\sigma}-\frac{2}{n-2}(
g_{\mu[\rho}R_{\sigma]\nu}-g_{\nu[\rho}R_{\sigma]\mu})+\frac{2}{(n-1)(n-2)}R
g_{\mu[\rho}g_{\sigma]\nu},\label{wten}
\end{eqnarray}
where $n$ and $g_{\mu\nu}$ are the dimension and metric of the spacetime, and brackets around indices refers to the antisymmetric part. Therefore, the electrodynamics with Weyl corrections (\ref{LEM}) is a special kind of electromagnetic theory which contains a coupling between the gravitational and electromagnetic fields. It was found that the such kind of couplings between curvature tensor and Maxwell tensor could be obtained from a calculation in QED of the photon effective action from one-loop vacuum polarization on a curved background \cite{Drummond}. Moreover, the investigations also show that these couplings could exist near classical compact astrophysical objects with high mass density and strong gravitational field such as the supermassive black holes at the center of galaxies \cite{Dereli1,Solanki}. Recently, many efforts have been focus on studying the effects of Weyl correction on black hole physics. In Ref.\cite{Weyl1}, the authors studied the holographic conductivity and charge diffusion with Weyl correction in the anti-de Sitter
 spacetime and found that the correction breaks the universal
relation with the $U(1)$ central charge observed at leading order.
Moreover, the holographic superconductors with Weyl corrections are
also studied in \cite{Wu2011,Ma2011,Momeni,Roychowdhury}. It is found that
Weyl corrections modify the critical temperature at which holographic superconductors occur \cite{Wu2011} and changes the order of the phase
transition of the holographic superconductor \cite{Ma2011}. The effects of Weyl corrections on the phase transition between the holographic insulator and
superconductor has been also investigated in \cite{zhao2013}. Recently, we \cite{sb2013} studied the dynamical evolution of the
electromagnetic perturbation coupling to the Weyl tensor in the
Schwarzschild black hole spacetime and analyze the effect of the Weyl
corrections on the stability of the black hole.

 It is well known that the properties and structure of a charged black hole depend heavily on the electrodynamics of Maxwell field in the spacetime, which means that the corrections to the standard Einstein-Maxwell theory must bring some new features for the charged black hole.
 The main purpose of this paper is to investigate the charged black hole in the electromagnetic theory with Weyl corrections, and to probe how the Weyl corrections modify its properties and structure.

The paper is organized as follows: in the following section we will
construct a static and spherically symmetric solution of a black hole with small Weyl corrections, and then study the effect of the Weyl coupling parameter $\alpha$ on the black hole. In Sec.III, we obtain a rotating charged black hole with small Weyl corrections by the method of complex coordinate
transformation \cite{JNT} and study the change of the spatial topology of the event horizons and the infinite redshift surface originating from the Weyl corrections.  We end the paper with a summary.

\section{A static and spherically symmetric charged black hole with Weyl correction}

Let us now first study a static and spherically symmetric charged black hole with Weyl correction. The action for the gravity system with the coupling between  electromagnetic field and Weyl tensor has a form
\begin{eqnarray}
S=\int d^4x \sqrt{-g}\bigg[R-\frac{1}{4}F_{\mu\nu}F^{\mu\nu}+\alpha
C^{\mu\nu\rho\sigma}F_{\mu\nu}F_{\rho\sigma}\bigg].\label{acts}
\end{eqnarray}
Adopting to Schwarzshild coordinates, the line element for a static spherically
symmetric spacetime can be put in the form
\begin{eqnarray}
ds^2&=&f(r)dt^2-\frac{1}{f(r)}dr^2-R(r)(d\theta^2+\sin^2{\theta}d\phi^2),
\label{m1}
\end{eqnarray}
where the metric coefficients $f(r)$ and $R(r)$ are functions of polar coordinate $r$. Moreover, we assume that the
electromagnetic field inherits the static spherically symmetries, which means that the potential four-vetor of the electric field has the form
\begin{eqnarray}
A_{\mu}=(\phi(r),0,0,0).\label{Au}
\end{eqnarray}
Inserting Eqs.(\ref{wten}), (\ref{m1}) and (\ref{Au}) into the action (\ref{acts}) and varying the action with respect to  $f(r)$, $R(r)$ and $A_{\mu}$, one can three coupled equations of motion
\begin{eqnarray}
&&3\bigg(R'(r)^2-2R(r)R''(r)\bigg)+4\alpha\frac{d}{dr}\bigg[\phi'(r)R(r)\bigg(2R(r)\phi''(r)
+R'(r)\phi'(r)\bigg)\bigg] =0,\label{Em1}
\\&&
3\bigg[\phi'(r)^2R(r)^2+f(r)\bigg(R'(r)^2-2R(r)R''(r)\bigg)-2R(r)\bigg(R(r)f''(r)+R'(r)f'(r)
\bigg)\bigg]\nonumber\\
&&-4\alpha\bigg[2R(r)\phi''(r)^2\bigg(\phi'(r)f(r)\bigg)'+2R(r)R'(r) \bigg(f(r)\phi'(r)^2\bigg)'-\phi'(r)^2\bigg(R'(r)^2f(r)
+2f''(r)R(r)^2\bigg)\nonumber\\&&+2f(r)R(r)\phi'(r)\bigg(R(r)\phi'''(r)
+R''(r)\phi'(r)\bigg)\bigg]=0,\label{Em2}
\\&&\frac{d}{dr}\bigg\{\phi'(r)R(r)+\frac{4\phi'(r)\alpha}{3R(r)}
\bigg[f(r)\bigg(R'(r)^2-R(r)R''(r)\bigg)+R(r)\bigg(R'(r)f''(r)-f'(r)R'(r)\bigg)
-2R(r)\bigg]\bigg\}=0.\nonumber\\&&\label{Em3}
\end{eqnarray}
In order to obtain a solution of a black hole with Weyl correction, we must solve these three coupling equations. As $\alpha\rightarrow 0$, one can find that Equation (\ref{Em1}) is decoupled naturally and then the solution of Reissner-Nordstr\"{o} m black hole can be obtained. However, for the case with non-zero Weyl coupling constant $\alpha$,  we find that
the modified equations of motion (\ref{Em1})-(\ref{Em3}) are so complicated that it is difficult for us to obtain an analytical solution of black hole. Here, we limit ourselves to the case where the deviation of the coupling parameter $\alpha$ from zero is very small which is
physically justified for the weak Weyl correction. Then the terms containing the parameter $\alpha$ on the left-hand-side of the three equations above
can be regarded as perturbation. Using the perturbation theory, we have
\begin{eqnarray}
&&R(r)=R_0(r)+\alpha R_1(r)+\mathcal{O}(\alpha^2),\nonumber\\
&&f(r)=f_0(r)+\alpha f_1(r)+\mathcal{O}(\alpha^2),\nonumber\\
&&\phi(r)=\phi_0(r)+\alpha \phi_1(r)+\mathcal{O}(\alpha^2).\label{wer}
\end{eqnarray}
Substituting the variables (\ref{wer}) into the equations of motion (\ref{Em1})-(\ref{Em3}), we can obtain a series of perturbational equations.
Obviously, the usual Reissner-Nordstr\"{o} m black hole is a solution of the zeroth order equations, which means that
\begin{eqnarray}
R_0(r)=r^2,~~~~~~ f_0(r)=1-\frac{2M}{r}+\frac{q^2}{r^2},~~~~~~
\phi_0(r)=\frac{q}{r}.\label{RN}
\end{eqnarray}
\begin{figure}
\begin{center}
\includegraphics[width=7cm]{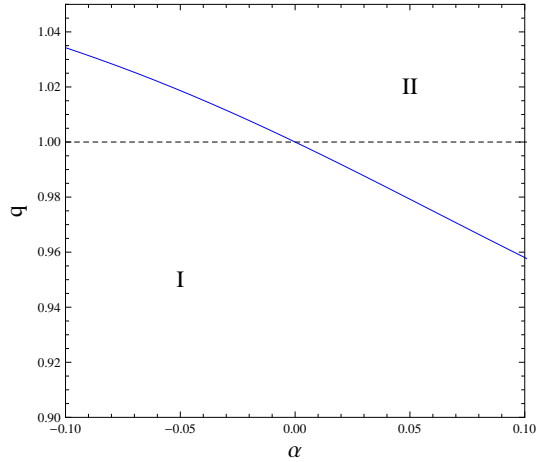}
\caption{In the $\alpha$-$q$ plane, the event horizon is existed only the region $I$ for the charged black hole with Weyl corrections (\ref{metr1}). In the region $II$, there does not exist any horizon and the central object is a naked singularity. The solid line marks the
upper limit on the charge $q$ as a function of the coupling parameter $\alpha$. The dashed line corresponds to the
upper limit on the charge $q$ in the Reissner-Nordstr\"{o} m black hole. Here, we set $M=1$.}
\end{center}
\end{figure}
Solving the first order equation, we obtain
\begin{eqnarray}
&&R(r)=r^2+\frac{4\alpha q^2}{9r^2}\nonumber\\
&&f(r)=1-\frac{2M}{r}+\frac{q^2}{r^2}-\frac{4\alpha q^2}{3r^4}\bigg(1-\frac{10M}{3r}+\frac{26q^2}{15r^2}\bigg),\nonumber\\
&&\phi(r)=\frac{q}{r}+\frac{\alpha q}{r^3}\bigg(\frac{M}{r}-\frac{37q^2}{45r^2}\bigg).\label{metr1}
\end{eqnarray}
Obviously, the metric coefficients and the static electric potential depend on the coupling parameter $\alpha$, which means that Weyl corrections affect the behavior of the electric field and the properties of the charged black hole in this case. Especially, we find that the static electric potential $\phi(r)$ depends also on the black hole parameter $M$, which is different from that in the usual Reissner-Nordstr\"{o} m black hole in which the static electric potential $\phi(r)$ depends only on the charge $q$. It is understandable because the Weyl coupling in here is a kind of coupling between the gravitational and electromagnetic fields.

The radius of black hole horizon is located at where $f(r)=0$ for the charged black hole with Weyl correction (\ref{metr1}). However, the equation $f(r)=0$ in this case could have more than two real roots. Considered that we here focus only on the weak Weyl correction, it is reasonable to regard the roots near those in the case of Reissner-Nordstr\"{o} m black hole as the radius of black hole horizon and to abandon other one as the extraneous roots of the equation. In this way, we can single out the roots corresponding to the radius of black hole horizons and probe the effects of Weyl corrections on the horizons. In figure (1), we delineate the region $I$ in the parameter
space $(\alpha, q)$, within which the event horizon is existed for the spacetime (\ref{metr1}). The solid line marks the
upper limit on the charge $q$ as a function of the coupling parameter $\alpha$,
for which the event horizon is still existed. The
region $II$ corresponds to the part of the parameter
space, where there does not exist any horizon and the central object is a naked singularity. From figure (1), we find that with Weyl corrections the allowed range of $q$ increases for the negative $\alpha$ and decreases for the positive $\alpha$.
In figure (2), we plot the effects of Weyl corrections on the inner and  outer horizons of the black hole as the parameters $\alpha$ and $q$ lie in the region $I$. It is shown that
with the increase of the Weyl corrections, the radius of outer horizon $r_+$ decreases, but the radius of inner horizon $r_-$ increases for $\alpha>0$, but   the situation is just the opposite for $\alpha<0$. The change of Hawking temperature $T_H$ of black hole with $\alpha$ is plotted in figure (3), which tells us that Hawking temperature decreases with $\alpha$. In the low energy limit, the luminosity of Hawking radiation of a spherically symmetric black hole (\ref{metr1}) can be approximated as $L\sim\frac{2\pi^3T^4_H}{15}(r^2_++\frac{4\alpha q^2}{9r^2_+})$. From figure (4), one can find that the luminosity of Hawking radiation also decreases with the Weyl coupling parameter $\alpha$.
\begin{figure}
\begin{center}
\includegraphics[width=7cm]{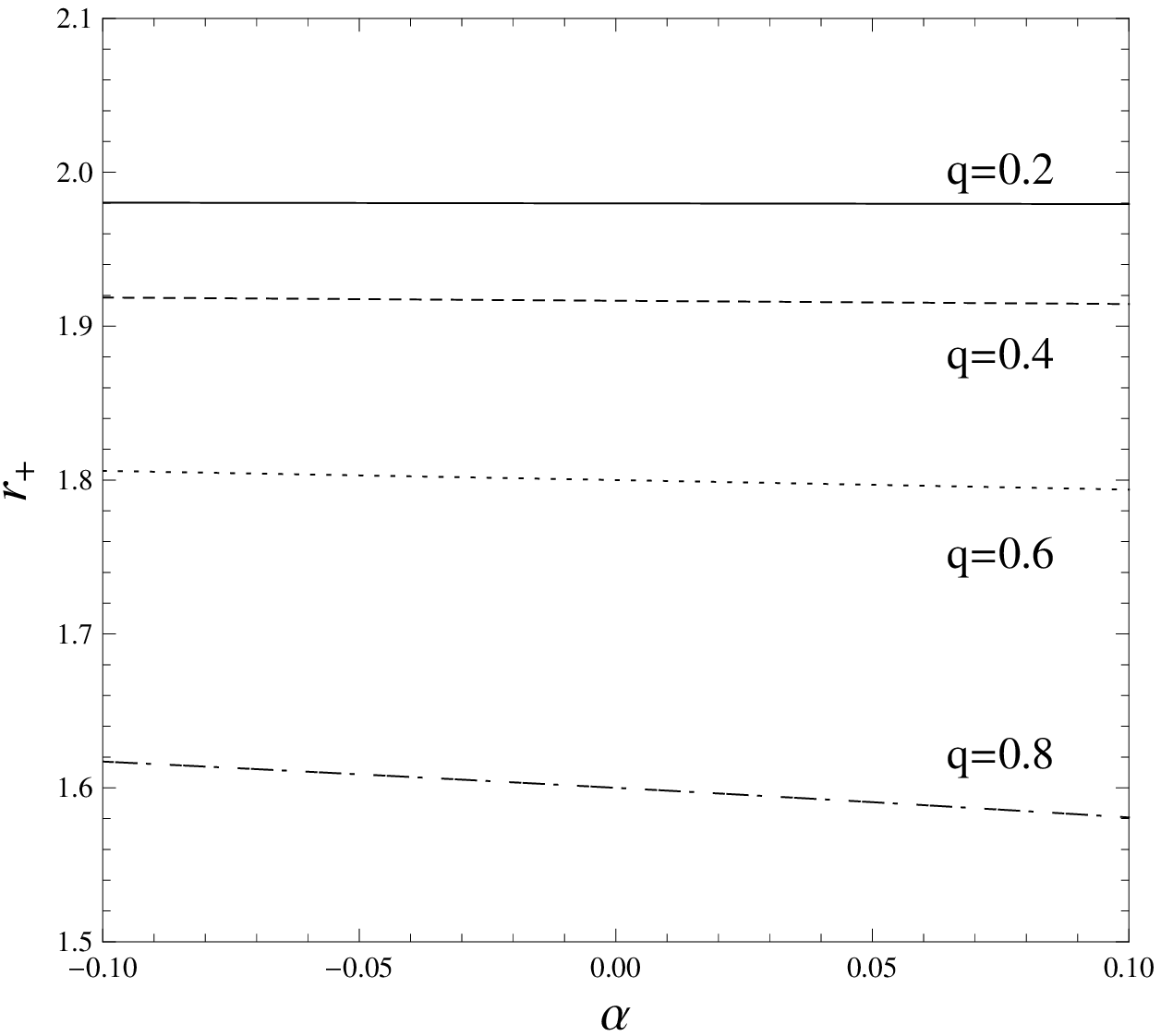}\includegraphics[width=7cm]{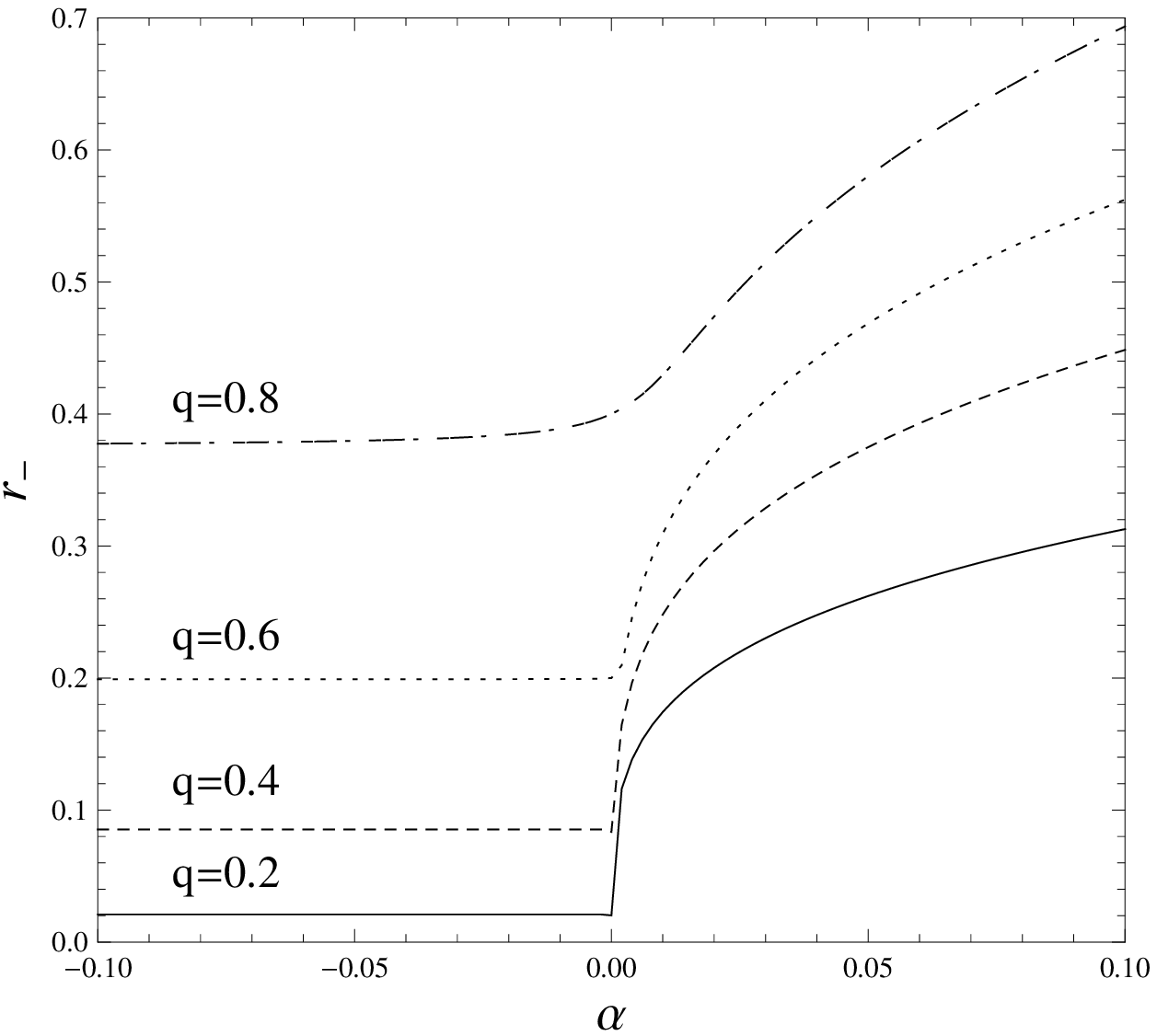}
\caption{The changes of the radius of horizon with Weyl coupling parameter $\alpha$ as the parameters $q$ and $\alpha$ lie in the region $I$. The left is for the outer horizon and the right is for the inner horizon. Here we set $M=1$. }
\end{center}
\end{figure}
\begin{figure}
\begin{center}
\includegraphics[width=7cm]{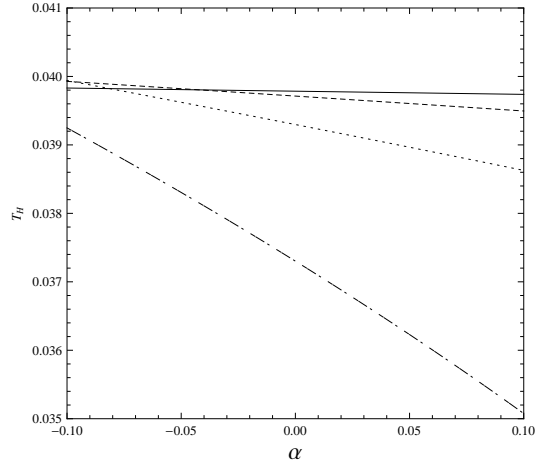}
\caption{The change of Hawking temperature $T_H$ of black hole with Weyl coupling parameter $\alpha$. Here we set $M=1$. }
\end{center}
\end{figure}
\begin{figure}
\begin{center}
\includegraphics[width=7cm]{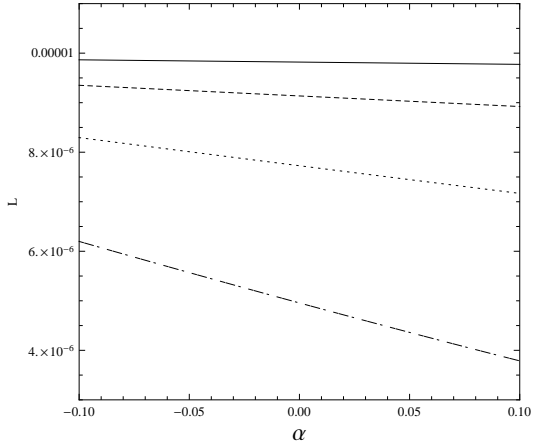}
\caption{The change of luminosity of Hawking radiation with  Weyl coupling parameter $\alpha$. Here we set $M=1$.}
\end{center}
\end{figure}
In a word, with the increase of the deviation from the Reissner-Nordstr\"{o} m metric, the effects of Wely corrections on the properties of the black hole in the case  $\alpha>0$ is different from that in the case  $\alpha<0$.

\section{A rotating charged black hole with small Weyl corrections}

In this section, we will study a rotating charged black hole with Weyl corrections. In the presence of the coupling between the Maxwell field and the Weyl tensor, the field equations of gravity for a rotating black hole are complicated. Even if in the case with small Weyl corrections, we find it is still very difficult to find exact analytical solution of a rotating black hole by usual perturbational method with some calculation softwares including Maple and Mathematica. Considering that the spacetime described by the action (\ref{acts}) is an electrovacuum solution, we can apply the method of complex coordinate
transformation discovered by Newman and Janis \cite{JNT} to construct a rotating black hole with the previous static black hole solution with small Weyl corrections (\ref{metr1}) and then study the properties of the black hole spacetime. Although our method in this section is different from the perturbational method used in the previous section, the solution is true because in the action (3) the only non-gravitational mass-energy present is the field energy of an electromagnetic field and the spacetime described by the action (3) is electro-vacuum and the approach of Newman and Janis is substantiated for vacuum or electr-ovacuum \cite{JNT1}. Moreover, in the following calculation, we neglect the terms of the order $\mathcal{O}(\alpha^2)$ and the higher-order terms to keep the consistency with the small Wey corrections just considered in the section II.

Introducing the new variable $u$ defined by
\begin{eqnarray}
u=t-\int \frac{dr}{f(r)},
\end{eqnarray}
one can rewrite the metric (\ref{metr1}) as
\begin{eqnarray}
ds^2&=&f(r)du^2+2dudr-R(r)^2(d\theta^2+\sin^2{\theta}
d\phi^2). \label{metric3}
\end{eqnarray}
The inverse of the above metric can be expressed as
\begin{eqnarray}
g^{\mu\nu}=-l^{\mu}n^{\nu}-l^{\nu}n^{\mu}+m^{\mu}\bar{m}^{\nu}
+m^{\nu}\bar{m}^{\mu},
\end{eqnarray}
with the null tetrad vectors
\begin{eqnarray}
l^{\mu}&=&\delta^{\mu}_1,\\
n^{\mu}&=&\delta^{\mu}_0-\frac{1}{2}\bigg[1-\frac{2M}{r}+\frac{q^2}{r^2}-\frac{4\alpha q^2}{3r^4}\bigg(1-\frac{10M}{3r}+\frac{26q^2}{15r^2}\bigg)\bigg]\delta^{\mu}_1,\\
m^{\mu}&=&\frac{1}{\sqrt{2R(r)}}\bigg(\delta^{\mu}_{2}
+\frac{i}{\sin{\theta}}\delta^{\mu}_{3}\bigg),\\
\bar{m}^{\mu}&=&\frac{1}{\sqrt{2R(r)}}\bigg(\delta^{\mu}_{2}
-\frac{i}{\sin{\theta}}\delta^{\mu}_{3}\bigg).
\end{eqnarray}
Now we regard the radius $r$ as a complex variable
and then rewrite the null tetrad in the form
\begin{eqnarray}
l^{\mu}&=&\delta^{\mu}_1,\\
n^{\mu}&=&\delta^{\mu}_0-\frac{1}{2}\bigg\{1-M\bigg(\frac{1}{r}+
\frac{1}{\bar{r}}\bigg)+\frac{q^2}{r\bar{r}}-\frac{4\alpha q^2}{3r^2\bar{r}^2}\bigg[1-\frac{5M}{3}\bigg(\frac{1}{r}+
\frac{1}{\bar{r}}\bigg)+\frac{26q^2}{15r\bar{r}}\bigg]\bigg\}
\delta^{\mu}_1,\\
m^{\mu}&=&\frac{1}{\sqrt{2R(r\bar{r})}}\bigg(\delta^{\mu}_{2}
+\frac{i}{\sin{\theta}}\delta^{\mu}_{3}\bigg),\\
\bar{m}^{\mu}&=&\frac{1}{\sqrt{2R(r\bar{r})}}\bigg(\delta^{\mu}_{2}
-\frac{i}{\sin{\theta}}\delta^{\mu}_{3}\bigg),
\end{eqnarray}
where $\bar{r}$ is the complex conjugate of $r$. As in Ref.\cite{JNT},
we can perform a complex coordinate transformation
\begin{eqnarray}
u'&=&u-ia\cos\theta,\\
r'&=&r+ia\cos\theta,\\
\theta'&=&\theta,~~~\phi'=\phi,
\end{eqnarray}
and find that the tetrad is transformed as
\begin{eqnarray}
l'^{\mu}&=&\delta^{\mu}_1,\\
n'^{\mu}&=&\delta^{\mu}_0-\frac{1}{2}\bigg[1-
\frac{2Mr'-q^2}{r'^2+a^2\cos^2{\theta}}-\frac{4\alpha q^2}{3(r'^2+a^2\cos^2{\theta})^2}\bigg(1-
\frac{50Mr'-26q^2}{15(r'^2+a^2\cos^2{\theta})}\bigg)
\bigg]\delta^{\mu}_1,\\
m'^{\mu}&=&\frac{1}{\sqrt{2R(r')}}\bigg[ia\sin\theta(\delta^{\mu}_0-
\delta^{\mu}_1)+\delta^{\mu}_{2}+\frac{i}{\sin{\theta}}\delta^{\mu}_{3}\bigg],\\
\bar{m}'^{\mu}&=&\frac{1}{\sqrt{2R(r')}}\bigg[-ia\sin\theta(\delta^{\mu}_0-
\delta^{\mu}_1)+\delta^{\mu}_{2}-\frac{i}{\sin{\theta}}\delta^{\mu}_{3}\bigg].
\end{eqnarray}
With help of this new tetrad, the metric of a rotating charge black hole with Weyl corrections can be described by
\begin{eqnarray}
g'^{\mu\nu}=-l'^{\mu}n'^{\nu}-l'^{\nu}n'^{\mu}+m'^{\mu}\bar{m}'^{\nu}
+m'^{\nu}\bar{m}'^{\mu}.\label{metric4}
\end{eqnarray}
In the coordinates ($u'$, $r'$, $\theta'$, $\phi'$), the covariant components of the metric (\ref{metric4}) can be expressed as
\begin{eqnarray}
g'_{00}&=&\frac{F(r',\theta')}{\Sigma(r',\theta')},~~~~~~~~~~~~
g'_{01}=1,~~~~~~~~~~~~g'_{13}=-a\sin^2\theta',\\
g'_{22}&=&-\Sigma_1(r',\theta'),~~~~~~~~~~~~
g'_{03}=\bigg[1-\frac{F(r',\theta')}{\Sigma(r',\theta')}\bigg]a\sin^2\theta',\\
g'_{33}&=&-\frac{\sin^2\theta'}{\Sigma(r',\theta')}\bigg[\Sigma(r',\theta')\Sigma_1(r',\theta')
+a^2\sin^2\theta'\bigg(2\Sigma(r',\theta')-F(r',\theta')\bigg)\bigg],
\end{eqnarray}
with
\begin{eqnarray}
\Sigma(r',\theta')&=&r'^2+a^2\cos^2\theta',\\
\Sigma_1(r',\theta')&=&r'^2+a^2\cos^2\theta'+\frac{4\alpha q^2}{9(r'^2+a^2\cos^2\theta')},\\
F(r',\theta')&=&r'^2+a^2\cos^2\theta'-2Mr'+q^2-\frac{4\alpha q^2}{3(r'^2+a^2\cos^2{\theta'})}\bigg(1-
\frac{50Mr'-26q^2}{15(r'^2+a^2\cos^2{\theta'})}\bigg).
\end{eqnarray}
In order to eliminate the elements $g'_{01}$ and $g'_{13}$, we must use a transformation \cite{TJo,FLM} to the coordinates ($u'$, $r'$, $\theta'$, $\phi'$) which is given by
\begin{eqnarray}
du'&=&dt-W(r',\theta')dr,\\
r'&=&r,~~~~\theta'=\theta,\\
d\phi'&=&d\phi-G(r',\theta')dr,
\end{eqnarray}
with
\begin{eqnarray}
W(r',\theta')&=&\frac{g'_{01}g'_{33}-g'_{03}g'_{13}}{g'_{00}g'_{33}-g'^2_{03}}
=\frac{\Sigma(r',\theta')[\Sigma_1(r',\theta')+a^2\sin^2\theta']}{
F(r',\theta')\Sigma_1(r',\theta')+\Sigma(r',\theta')a^2\sin^2\theta'},\\
G(r',\theta')&=&\frac{g'_{00}g'_{13}-g'_{01}g'_{03}}{g'_{00}g'_{33}-g'^2_{03}}
=\frac{a\Sigma(r',\theta')}{F(r',\theta')\Sigma_1(r',\theta')+\Sigma(r',\theta')a^2\sin^2\theta'}.
\end{eqnarray}
And then the metric for a rotating charged black hole with Weyl corrections reads
\begin{eqnarray}
ds^2&=&\frac{F(r,\theta)}{\Sigma(r,\theta)}dt^2
+2\bigg[1-\frac{F(r,\theta)}{\Sigma(r,\theta)}\bigg]a\sin^2\theta dtd\phi
-\frac{\Sigma(r,\theta)\Sigma_1(r,\theta)dr^2}{F(r,\theta)\Sigma_1(r,\theta)
+a^2\sin^2\theta\Sigma(r,\theta)}
-\Sigma_1(r,\theta)d\theta^2\nonumber\\&-&\frac{\sin^2\theta}{\Sigma(r,\theta)}
\bigg[\Sigma(r,\theta)\Sigma_1(r,\theta)
+a^2\sin^2\theta\bigg(2\Sigma(r,\theta)-F(r,\theta)\bigg)\bigg]d\phi^2, \label{metric5}
\end{eqnarray}
with
\begin{eqnarray}
\Sigma(r,\theta)&=&r^2+a^2\cos^2\theta,\\
\Sigma_1(r,\theta)&=&r^2+a^2\cos^2\theta+\frac{4\alpha q^2}{9(r^2+a^2\cos^2\theta)},\\
F(r,\theta)&=&r^2+a^2\cos^2\theta-2Mr+q^2-\frac{4\alpha q^2}{3(r^2+a^2\cos^2{\theta})}\bigg(1-
\frac{50Mr-26q^2}{15(r^2+a^2\cos^2{\theta})}\bigg).
\end{eqnarray}
Obviously, the above metric can be reduced to the Kerr-Newman metric in Boyer-Lindquist coordinates as $\alpha=0$. When the rotation parameter $a$ vanishes, one can get the previous solution of a static and spherically symmetric black hole with Weyl correction (\ref{metr1}).

The mass and angular momentum of the rotating black hole with Weyl corrections (\ref{metric5}) can be calculated by the quasi-local formalism of the Brown and York \cite{BY}, which is extensively applied to various rotating black holes \cite{STZ,RBM,Sheykhi1,Hendi,LBS}. From the quasi-local formalism, one can find that the finite stress-energy tensor is defined as
\begin{eqnarray}
T^{ij}=\frac{1}{8\pi}[\Theta^{ij}-\Theta\gamma^{ij}],\label{T50}
\end{eqnarray}
where $\Theta$ is the trace of the extrinsic curvature $\Theta^{ij}$
of the boundary $\partial\mathcal{M}$ of the manifold $\mathcal{M}$, with the induced metric $\gamma_{ij}$.
In order to compute the angular
momentum of the spacetime, one can choose a spacelike surface $\mathfrak{B}$ in $\partial\mathcal{M}$ with the metric $\sigma_{ab}$ and decompose the boundary metric into the ADM form
\begin{eqnarray}
\gamma_{ij}dx^idx^j=-N^2dt^2+\sigma_{ab}(d\varphi^{a}+V^{a}dt)(d\varphi^{b}+V^{b}dt),
\end{eqnarray}
where the coordinates $\varphi^{a}$ are the angular variables parameterizing the hypersurface of constant $r$.  The quantities $N$ and $V^a$ are the lapse and shift functions respectively. If there is a Killing vector field $\xi$ on the boundary, one can find that the quasi-local conserved charge associated with the stress tensors can be defined by
\begin{eqnarray}
\mathcal{Q}(\xi)=\int_{\mathcal{B}}d^2\varphi\sqrt{\sigma}T_{ij}n^{i}\xi^{j}.
\end{eqnarray}
where $\sigma$ is the determinant of the metric $\sigma_{ab}$ , $\xi^{i}$ and $n^{i}$ are the Killing vector field and the
unit normal vector on the boundary $\mathcal{B}$, respectively.
For the boundary with timelike ($\xi=\frac{\partial}{\partial t}$) and
rotational ($\zeta=\frac{\partial}{\partial \varphi}$) Killing vector fields, one can write the quasi-local mass and angular momentum as in the forms
\begin{eqnarray}
\mathcal{M}&=&\int_{\mathcal{B}}d^2\varphi\sqrt{\sigma}T_{ij}n^{i}\xi^{j},\nonumber\\
J&=&\int_{\mathcal{B}}d^2\varphi\sqrt{\sigma}T_{ij}n^{i}\zeta^{j},\label{T53}
\end{eqnarray}
Combining the metric (\ref{metric5}) with Eqs.(\ref{T50})-(\ref{T53}), we obtain the quasi-local mass and angular momentum 
\begin{eqnarray}
\mathcal{M}&=&\lim_{r\rightarrow\infty}\bigg[M+\frac{M^2-q^2}{2r}
+\frac{3M^3-3Mq^2-2Ma^2}{6r^2}\nonumber\\&+&\frac{15M^4-18M^2q^2+3q^4-40M^2a^2-4a^2 q^2+90a^4+16\alpha q^2}{24r^3}+\mathcal{O}\bigg(\frac{1}{r^4}\bigg)\bigg]=M,\\
J&=&\lim_{r\rightarrow\infty}\bigg[Ma-\frac{2aq^2}{3r}
-\frac{2aq^2(a^2-10\alpha )}{15r^3}+\mathcal{O}\bigg(\frac{1}{r^4}\bigg)\bigg]=Ma.
\end{eqnarray}
Obviously, the quasi-local mass and angular momentum of the rotating black hole with Weyl corrections (\ref{metric5}) coincide with those in the usual Kerr-Newman black hole spacetime. The main reason is that these quasi-local mass and angular momentum are decided by the properties of the surface $\mathcal{B}$ at spatial infinity at where the effects of Weyl corrections are vanished in the dominant terms. This means that the Wely corrections in the Lagrangian density (\ref{LEM}) do not change the quasi-local mass and angular momentum of the black hole (\ref{metric5}).

The electric charge of the black hole with weyl correction (\ref{metric5}) can be obtained by calculating the flux of the electric field at infinity
\begin{eqnarray}
Q=\frac{1}{4\pi}\oint_{S_{\infty}}
\frac{\partial L_{EM}}{\partial F^2}F^{\mu\nu}dS_{\mu\nu}.\label{mq0}
\end{eqnarray}
Here $F^2=\frac{1}{4}F^{\mu\nu}F_{\mu\nu}$ and $dS_{\mu\nu}$ is the area element of a two-dimensional closed space-like surface at the spatial infinity $S_{\infty}$, which has a form
\begin{eqnarray}
dS_{\mu\nu}=\frac{1}{2}\sqrt{-g}\varepsilon_{\mu\nu\rho\sigma}dx^{\rho}\wedge
dx^{\sigma},~~~~~~~~\varepsilon_{tr\theta\phi}=1,
\end{eqnarray}
At spatial infinity $r\rightarrow\infty$, the potential four-vetor of the electric field of the black hole (\ref{metric5}) can be approximated as
\begin{eqnarray}
A_{t}&=&\frac{q}{r}-\frac{qa^2\cos^2{\theta}}{r^3}+\frac{\alpha Mq}{r^4}+\mathcal{O}\bigg(\frac{1}{r^5}\bigg),\nonumber\\
A_{\phi}&=&\frac{qa\sin^2{\theta}}{r}-\frac{qa^3\sin^2{\theta}\cos^2{\theta}}{r^3}
+\frac{\alpha Mqa\sin^2{\theta}}{r^4}+\mathcal{O}\bigg(\frac{1}{r^5}\bigg), \label{Au51}
\end{eqnarray}
Substituting it into the integral (\ref{mq0}), we can obtain
\begin{eqnarray}
Q=\lim_{r\rightarrow\infty}\bigg[q+\frac{qa^2}{3r^2}
+\frac{2Mq(a^2+18\alpha)}{3r^3} -\frac{q^3(3a^2+107\alpha)}{9r^4}
+\mathcal{O}\bigg(\frac{1}{r^5}\bigg)\bigg]=q.
\end{eqnarray}
It indicates that the electric charge of the rotating black hole with Weyl corrections is still $Q=q$ and the presence of Weyl corrections also does not affect the electric charge of the black hole.

\begin{figure}[ht]
\begin{center}
\includegraphics[width=5.5cm]{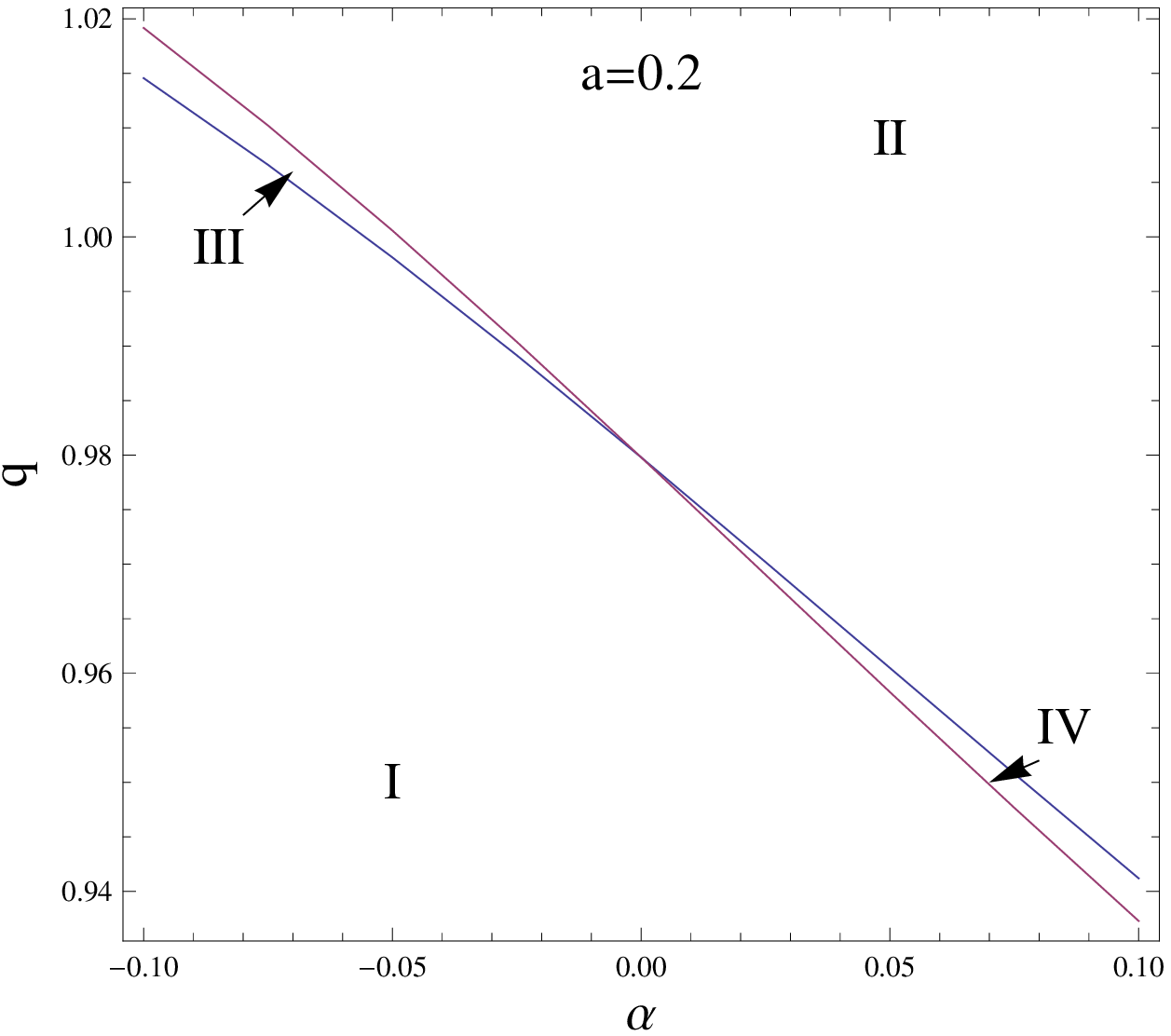}\includegraphics[width=5.5cm]{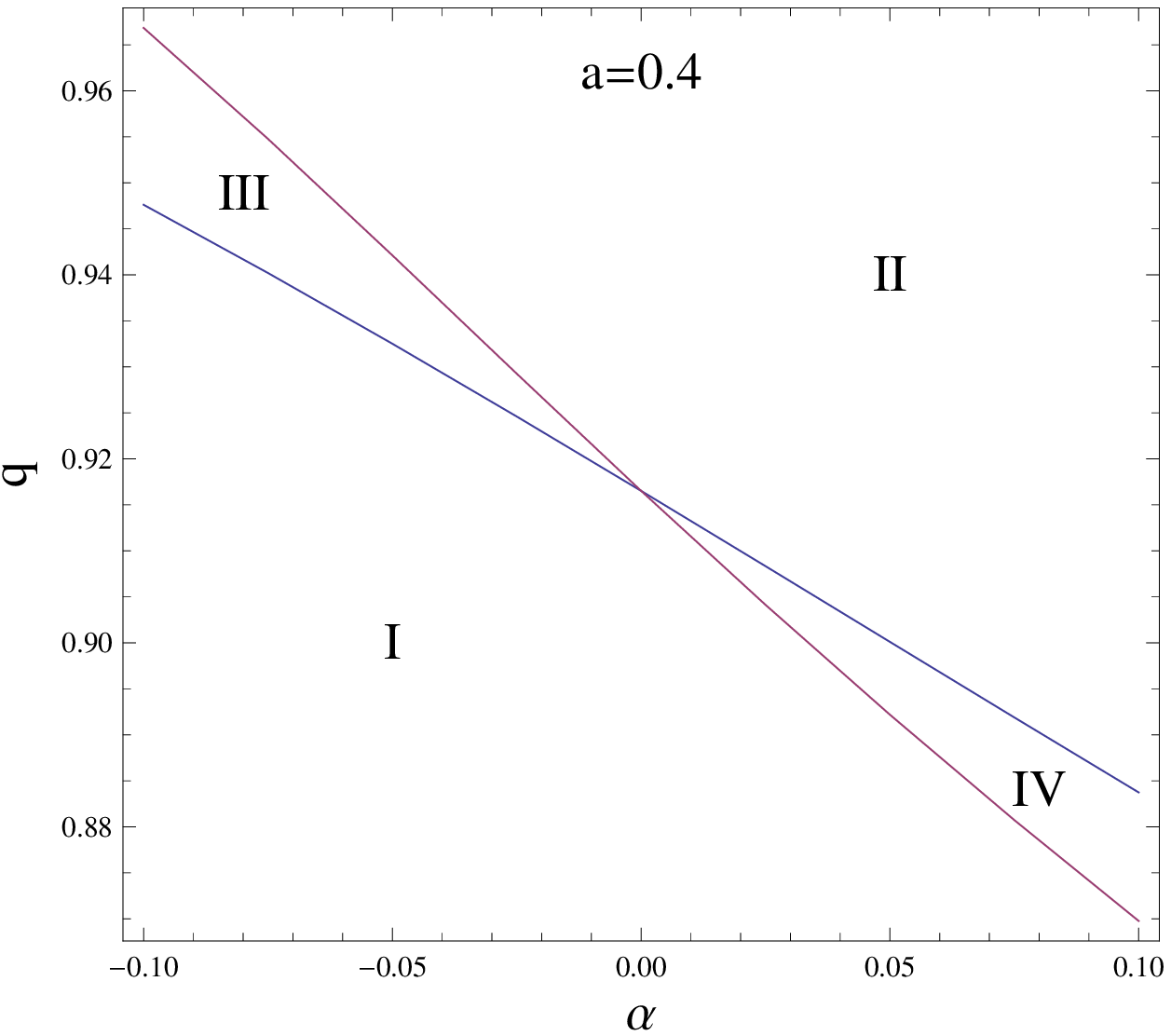}
\includegraphics[width=5.5cm]{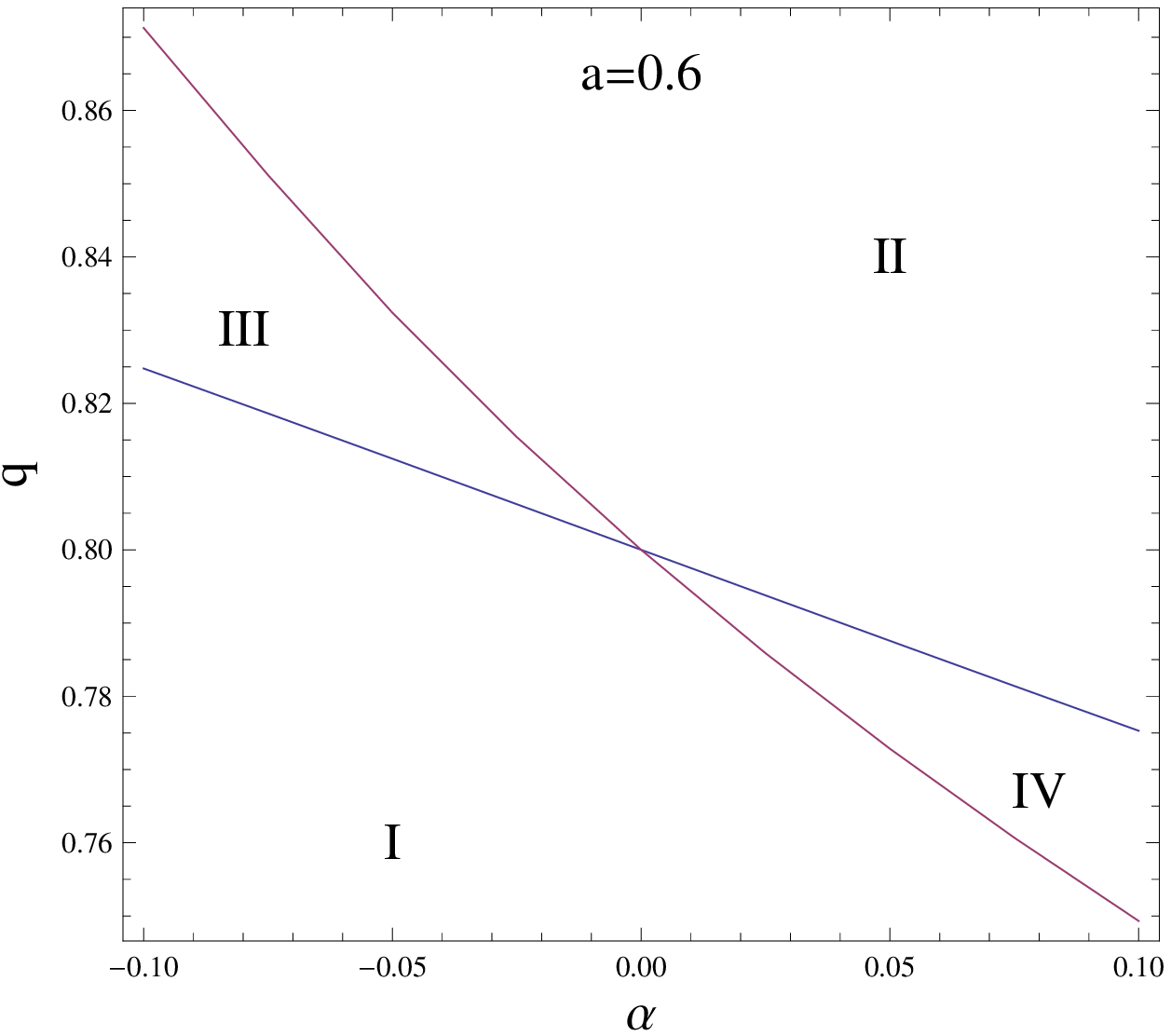}
\caption{In the $\alpha$-$q$ plane, the inner and outer horizons are topologically spherical surfaces without intersection in the region I. As the parameters lie in the region II, there does not exist any horizon for black hole with Weyl corrections.  The horizons merge into a closed toroidal surface in the region III and the horizons are disconnected in the region IV for a black hole with Weyl corrections. The panels from left to right correspond to the case $a=0.2, 0.4$ and $0.6$, respectively.  Here we set $M=1$.}
\end{center}
\end{figure}
\begin{figure}[ht]
\begin{center}
\includegraphics[width=6.6cm]{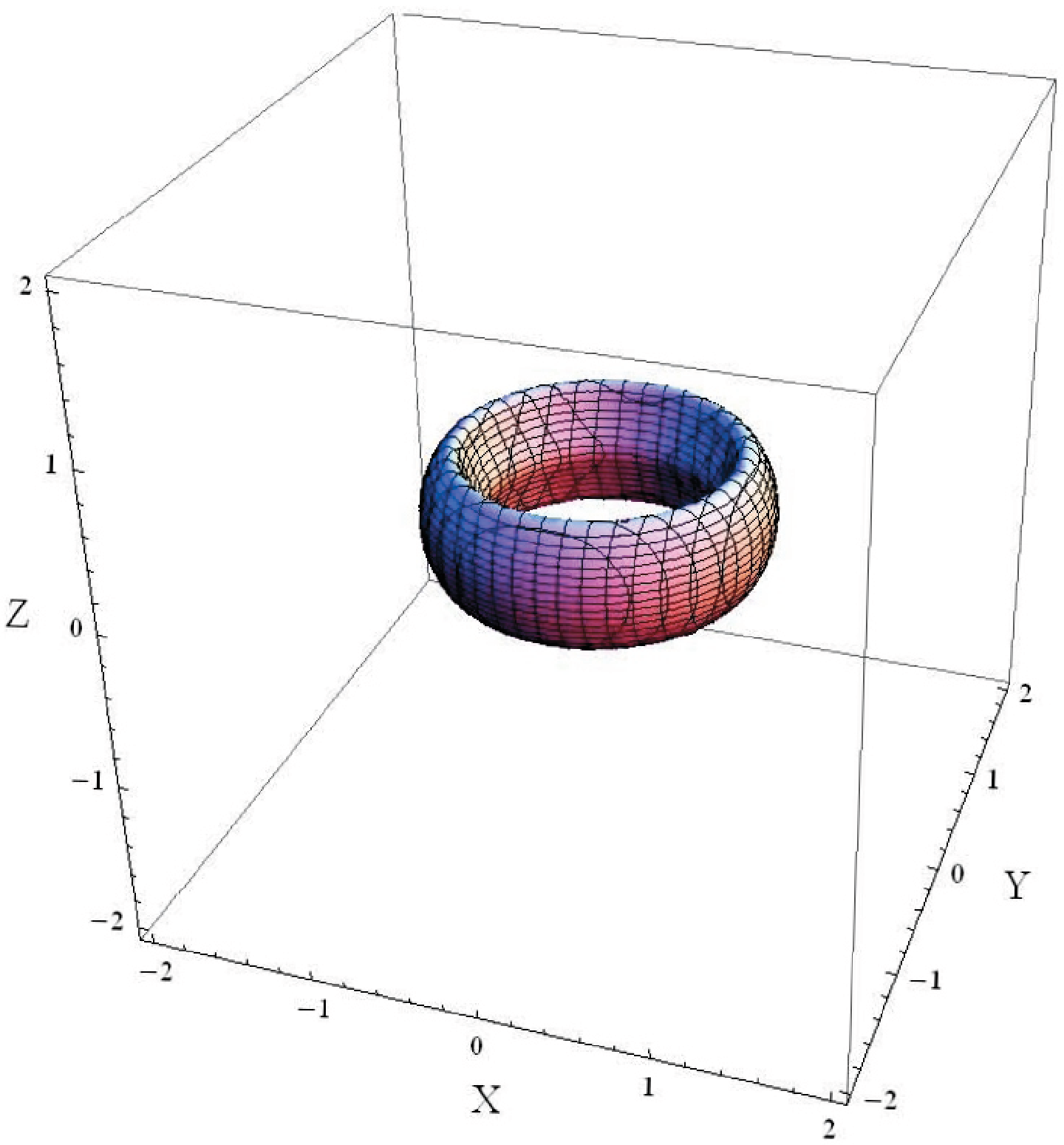}\includegraphics[width=7cm]{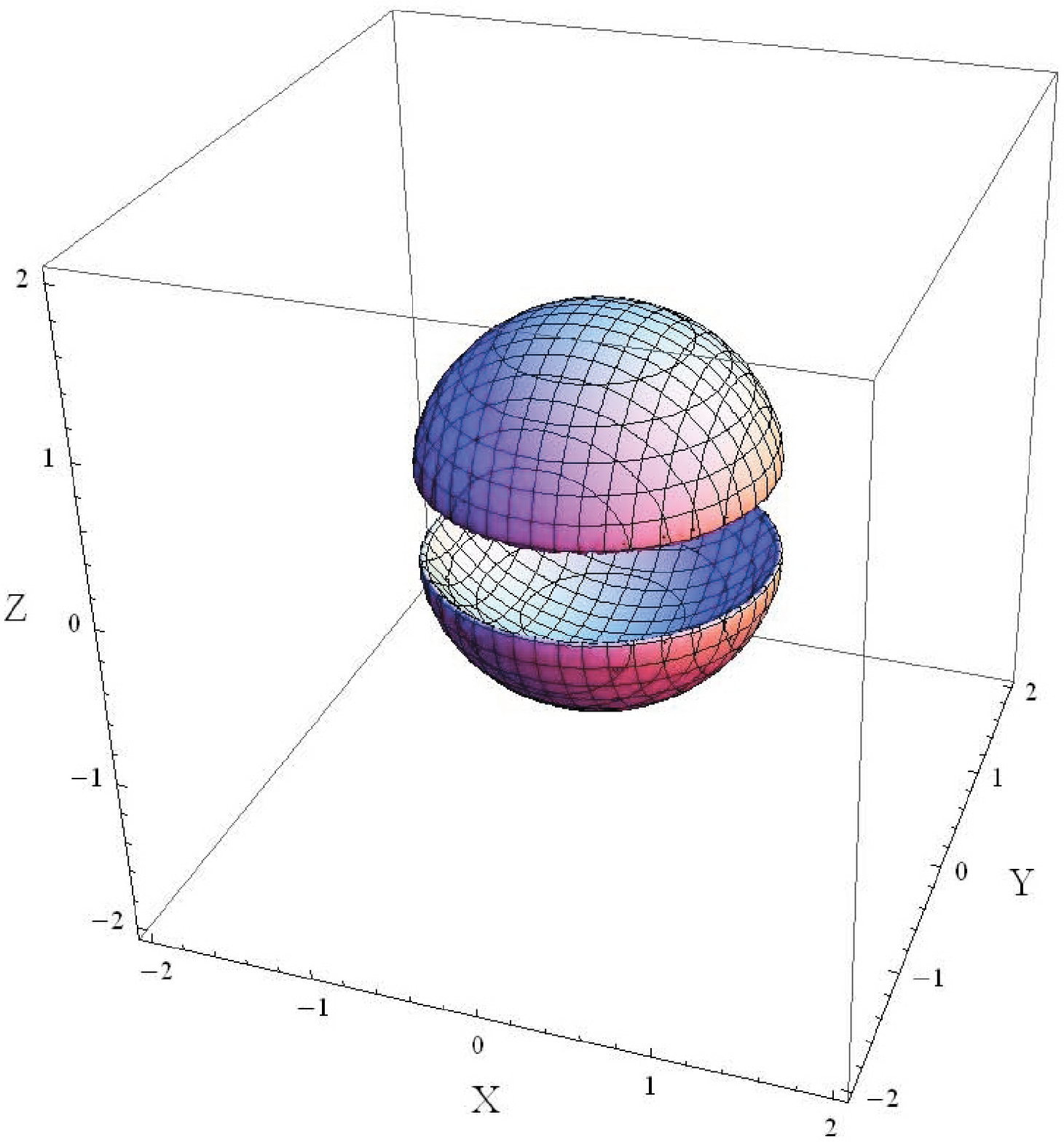}
\caption{Event horizons of a rotating charged black hole with Weyl corrections as the parameters lie in the region III or IV.  The left panel is for the black hole with the fixed parameters $a=0.6$,$\alpha=-0.075$ and $q=0.840$. The right panel is for the black hole with the fixed parameters $a=0.6$, $\alpha=0.075$ and $q=0.762$. Here we set $M=1$.}
\end{center}
\end{figure}

Let us now study the properties of such a rotating charged black hole with Weyl corrections (\ref{metric5}). The position of the black hole horizon is defined by equation
\begin{eqnarray}
g^2_{03}-g_{00}g_{33}=0,
\end{eqnarray}
i.e.,
\begin{eqnarray}
r^2-2Mr+a^2+q^2-\frac{4\alpha q^2}{45(r^2+a^2\cos^2\theta)}\bigg(10-
\frac{40M-21q^2}{r^2+a^2\cos^2\theta}\bigg)=0.
\end{eqnarray}
It indicates that for $\alpha\neq0$ the radius $r_{\pm}$ depends on the polar angle coordinate $\theta$, which is similar to those in the modified Kerr
metrics by the deformation parameter $\epsilon$ \cite{TJo}  or the polymeric function $P$ in loop quantum gravity \cite{FLM}. Furthermore, the position and the shape of horizons are defined by the parameters $M$, $a$, $\alpha$ and $q$.  For a rotating charged black hole with Weyl corrections, we find that the whole parameter space ($\alpha$-$q$) can be divided into four regions for fixed $a$ as shown in figure (5). In the region I, we find that both of the inner and outer horizons are topologically spherical surfaces and these two surfaces never cross each other. In the region II, there exist no horizons and the singularity is naked entirely. These properties of black hole are similar to those in the non-rotating black hole with Weyl corrections. However, we also find that when the parameters ($\alpha$, $q$) lie in  the region III, the outer horizon  coincides the inner horizon near the north and south poles and the horizons merge into a closed toroidal surface (see in the left panel of figure(6)). When the parameters ($\alpha$, $q$) lie in  the region IV, the parts of the outer and inner horizons in the northern hemisphere join together to form a new closed surface with spherical
topology. The similar case also occurs in the southern hemisphere, and then two new and disconnected horizons are formed, which is shown in the right panel in Fig. (6). These properties of black holes are not observed in the
non-rotating black hole with Weyl corrections. Moreover, we also note that
the value of the Weyl coupling parameter $\alpha$ is negative in the region III and is positive in the region IV, which means that the effects of the Weyl corrections with positive $\alpha$ on the black hole are quite different from those in the case with negative $\alpha$. It is easy for us to find that
properties of the rotating charged black hole with Weyl corrections (\ref{metr1}) are similar to the properties of the rotating non-Kerr black hole \cite{TJo,CBa}. The unique difference is that when the outer and inner horizons merge into a closed toroidal surface the singularity is naked in the rotating charged black hole with Weyl corrections, but it is enveloped by the toroidal surface in the rotating non-Kerr black hole.
With the increasing Weyl corrections, the value of the upper limit of $q$
in the regions I and III is increasing for $\alpha<0$, but the value of the upper limit of $q$ in the regions I and IV is decreasing for $\alpha>0$. Moreover, with the increase of the rotation parameter $a$, the range of the regions III and IV increases.
Furthermore, we find that as the parameters lie in the region I (see Fig.(5)) the radius of the outer horizon for a black hole with $\alpha>0$ is smaller than  that in the case with $\alpha<0$, which is consistent with those in a static and spherically symmetric black hole spacetime with Weyl corrections (\ref{metr1}).

The ergosphere is an important zone around a rotating black hole, which is bounded by the event horizon $r_+$ and the outer infinite redshift surface $r^+_{\infty}$.
The infinite redshift surface is determined by $g_{00}=0$, i.e.,
\begin{eqnarray}
r^2-2Mr+q^2+a^2\cos^2\theta-\frac{4\alpha q^2}{3(r^2+a^2\cos^2\theta)}\bigg(1-
\frac{50M-26q^2}{15(r^2+a^2\cos^2\theta)}\bigg)=0.
\end{eqnarray}
Similarly, the whole parameter space ($\alpha$-$q$) can be divided into three regions to study the properties of the infinite redshift surface for fixed $a$, which is shown in figure (7). We can find that the inner and outer infinite redshift surfaces are topologically spherical surfaces without intersection as the parameters $(\alpha, q)$ lie in the region I and there is no any infinite redshift surface as the parameters $(\alpha, q)$ lie in the region II. In the region III, the inner and outer infinite redshift surfaces merge into a new infinite redshift surface with toroidal topology around the original point, which is shown in figure (8). The new infinite redshift surface becomes more and more thin and looks like a disk as the rotation parameter $a$ increases.
\begin{figure}[ht]
\begin{center}
\includegraphics[width=5.5cm]{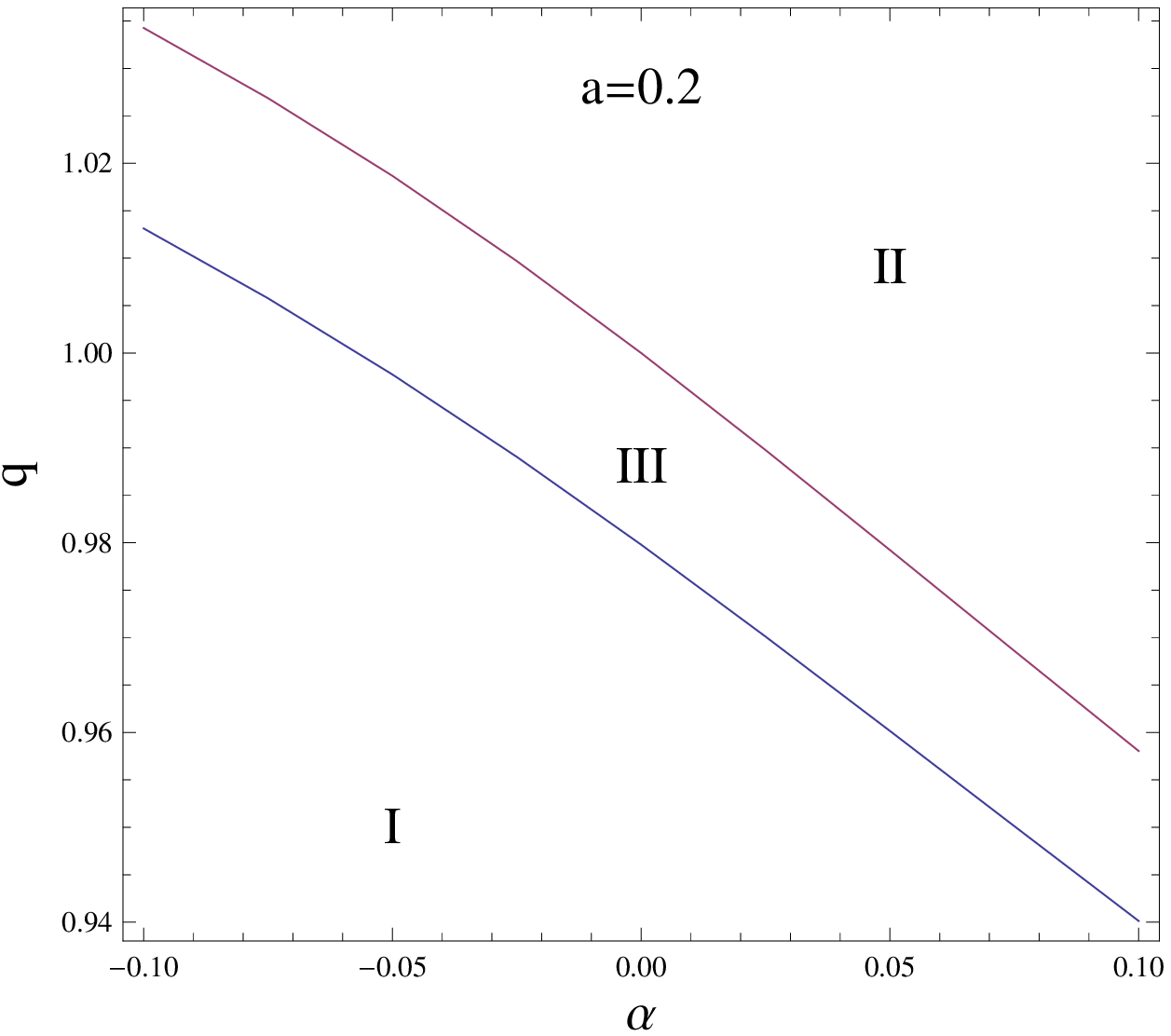}\includegraphics[width=5.5cm]{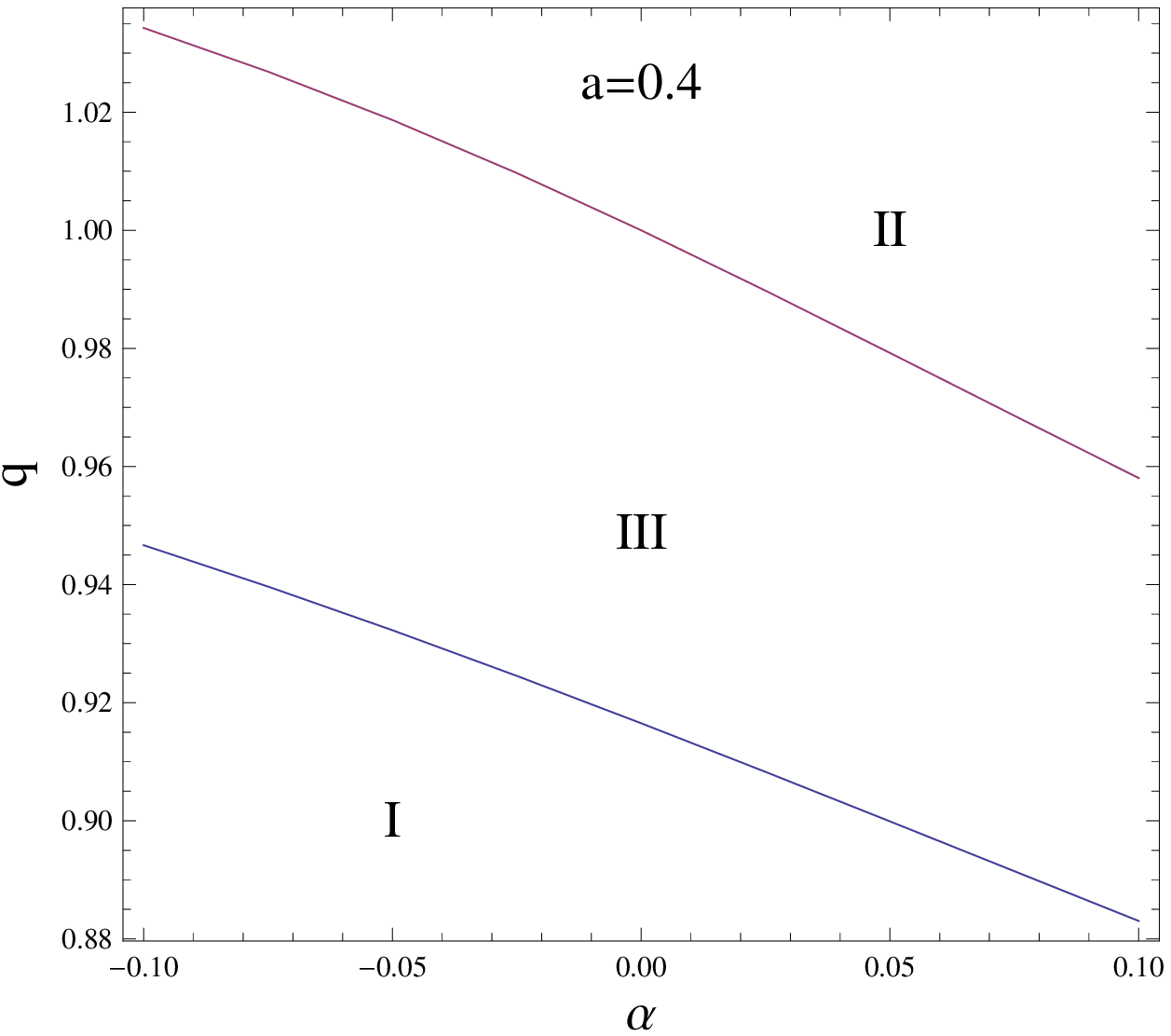}
\includegraphics[width=5.5cm]{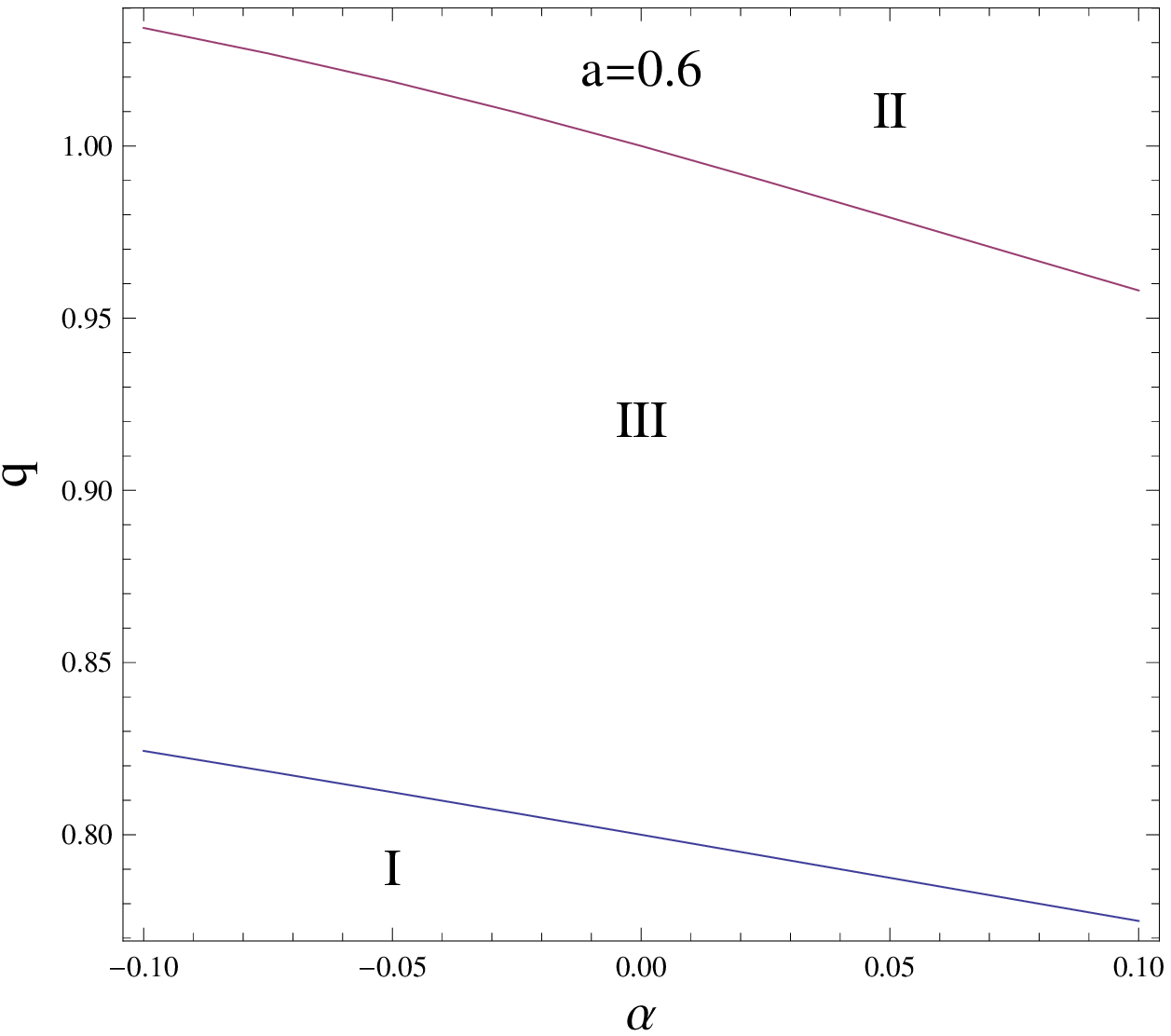}
\caption{In the $\alpha$-$q$ plane, the inner and outer infinite redshift surfaces are topologically spherical surfaces without intersection in the region I. As the parameters lie in the region II, there does not exist any infinite redshift surface for black hole with Weyl corrections.  The infinite redshift surfaces merge into a closed toroidal surface in the region III. The panels from left to right correspond to the case $a=0.2, 0.4$ and $0.6$, respectively.  Here we set $M=1$.}
\end{center}
\end{figure}
\begin{figure}[ht]
\begin{center}
\includegraphics[width=6.6cm]{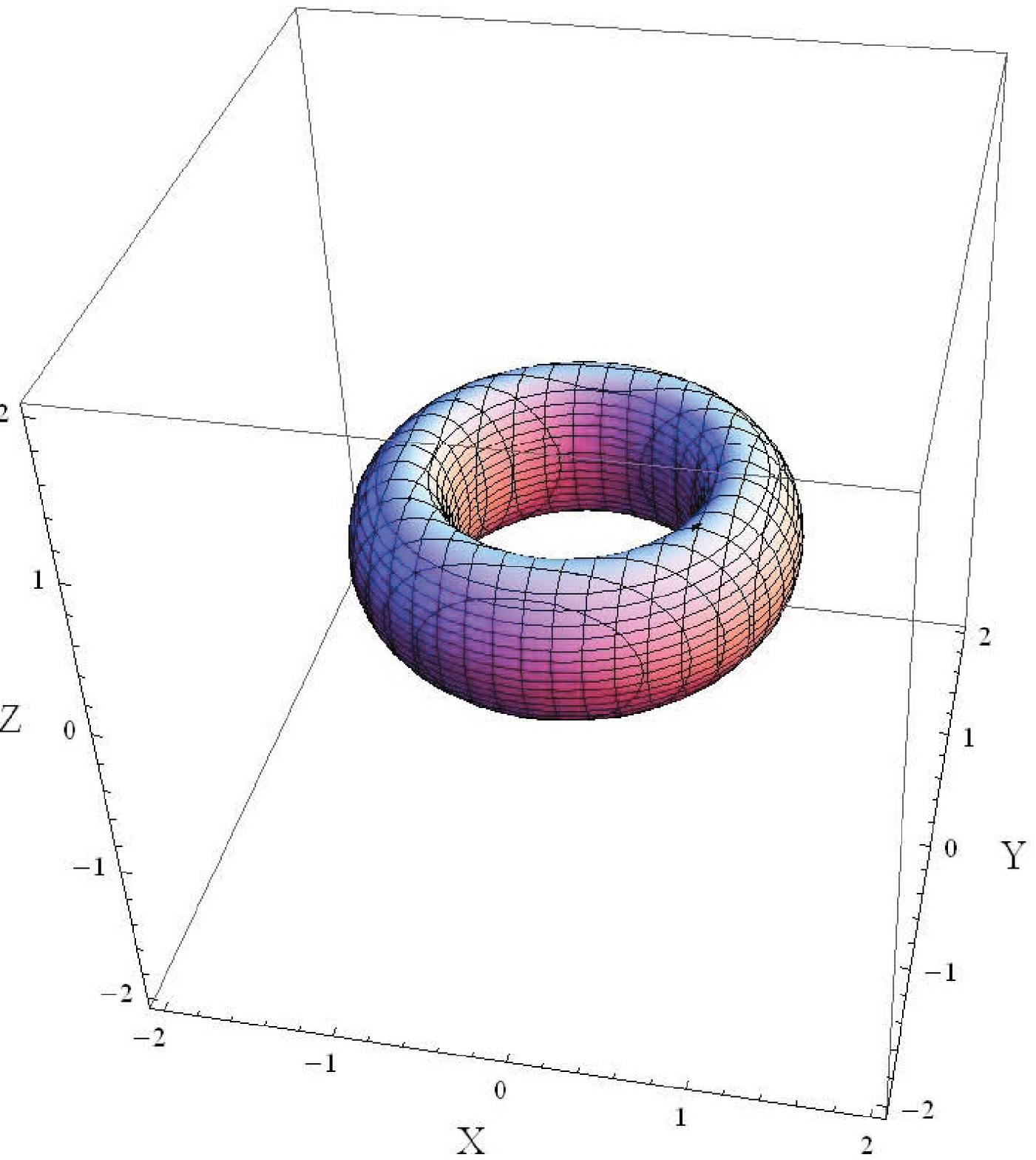}\includegraphics[width=7cm]{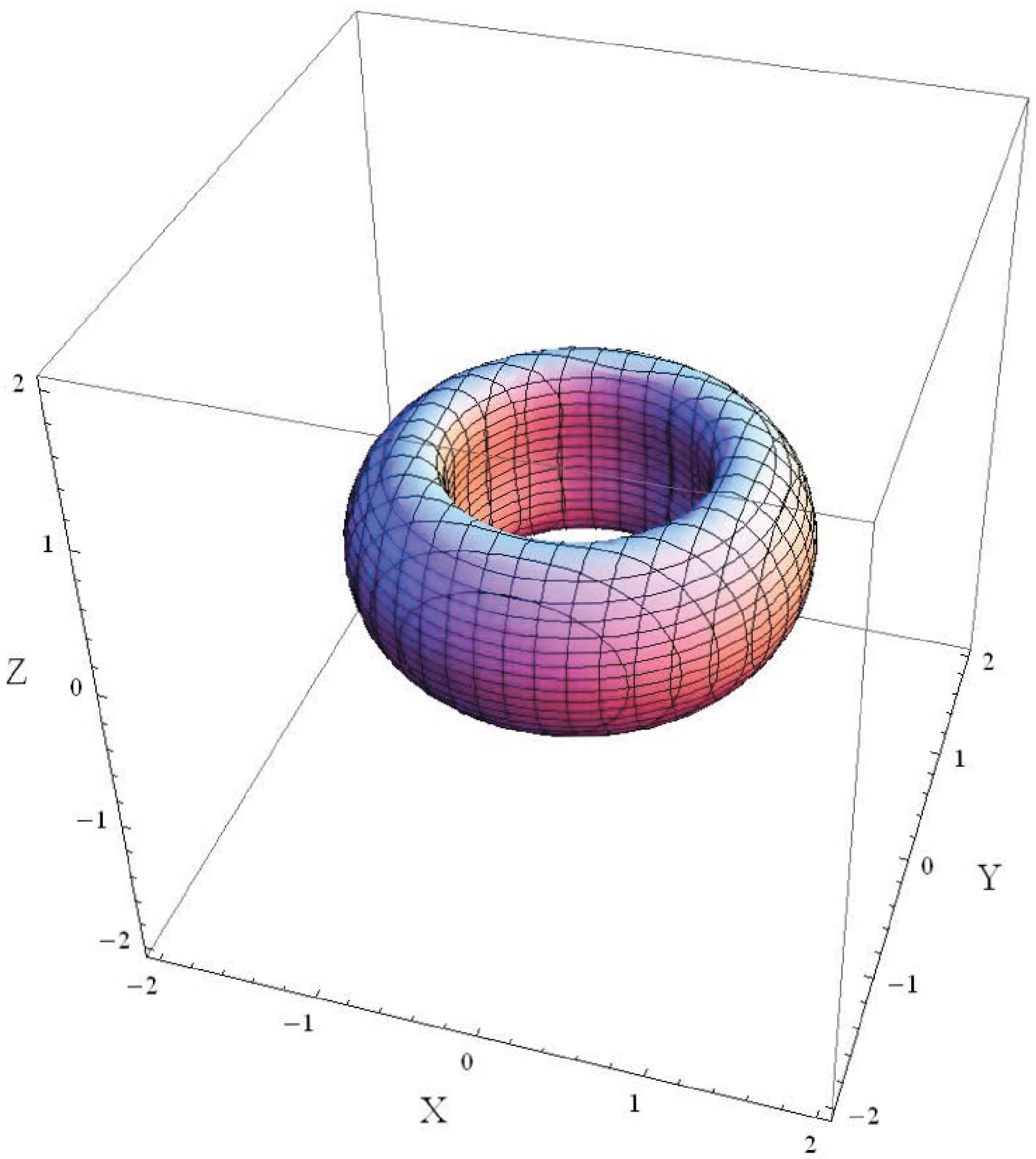}
\caption{The infinite redshift surfaces of a rotating charged black hole with Weyl corrections as the parameters lie in the region III.  The left panel is for the black hole with the fixed parameters $a=0.6$, $\alpha=-0.075$ and $q=0.98$. The right panel is for the black hole with the fixed parameters $a=0.6$, $\alpha=0.075$ and $q=0.91$. Here we set $M=1$..}
\end{center}
\end{figure}
\begin{figure}[ht]
\begin{center}
\includegraphics[width=5.5cm]{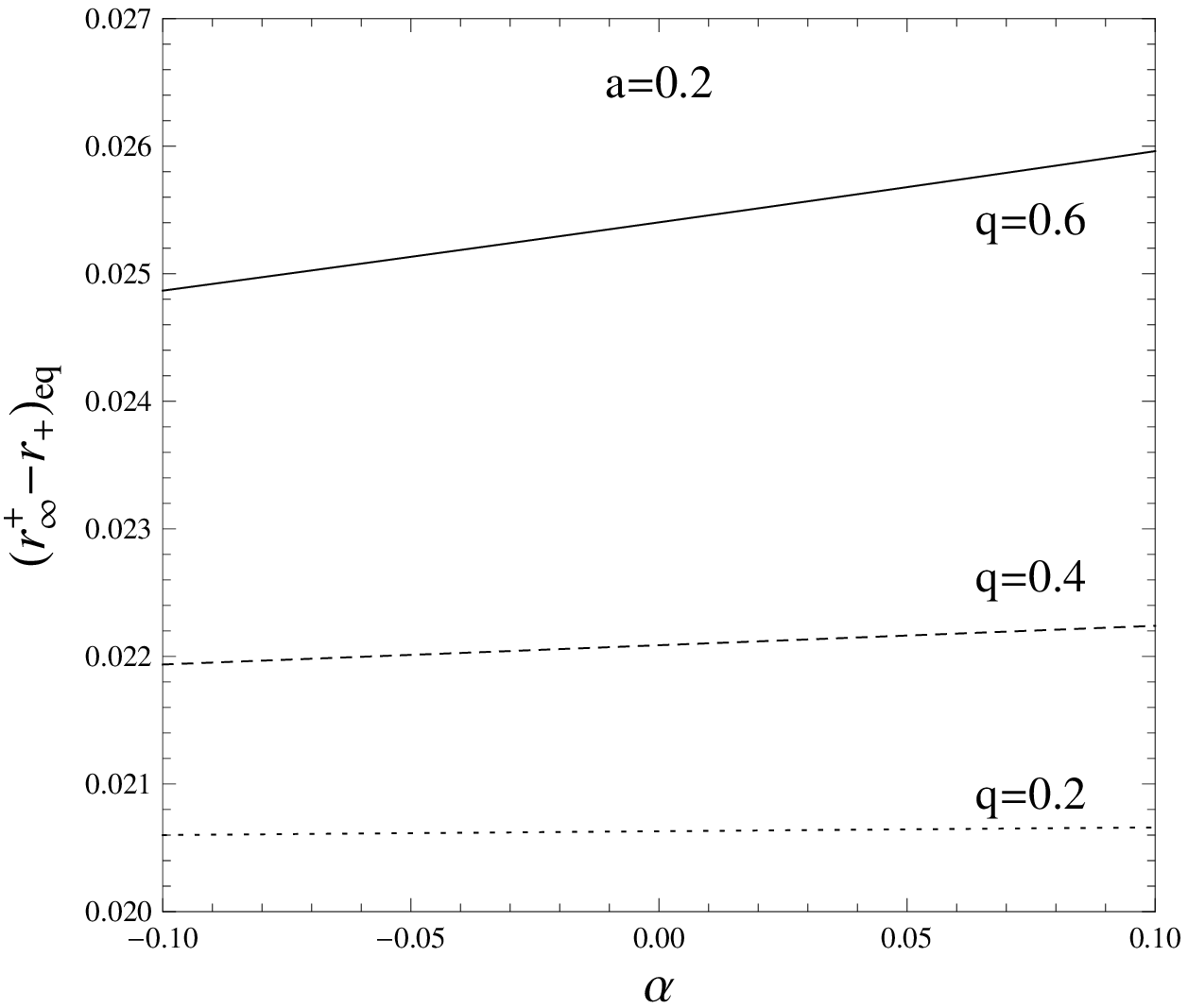}
\includegraphics[width=5.5cm]{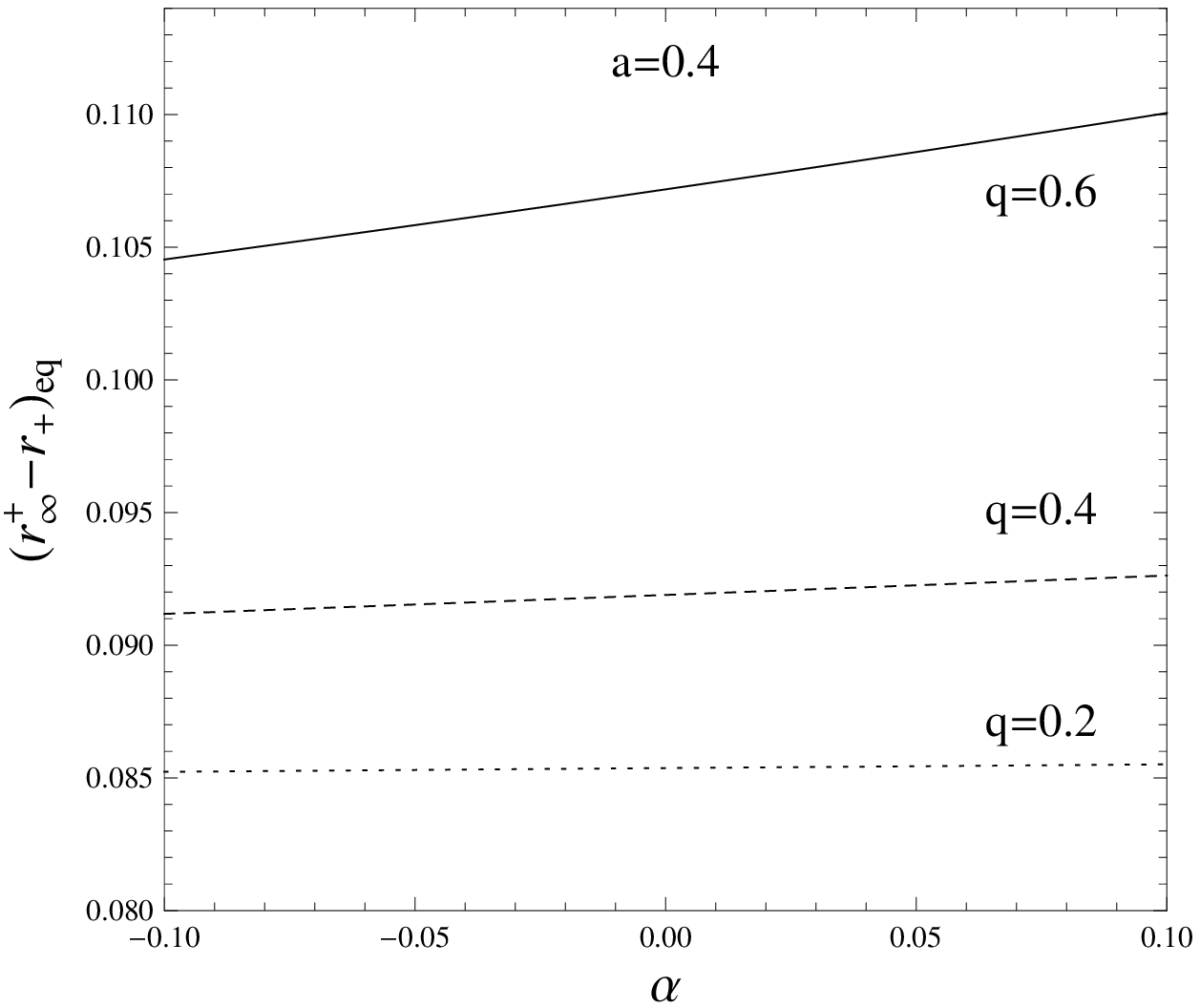}
\includegraphics[width=5.5cm]{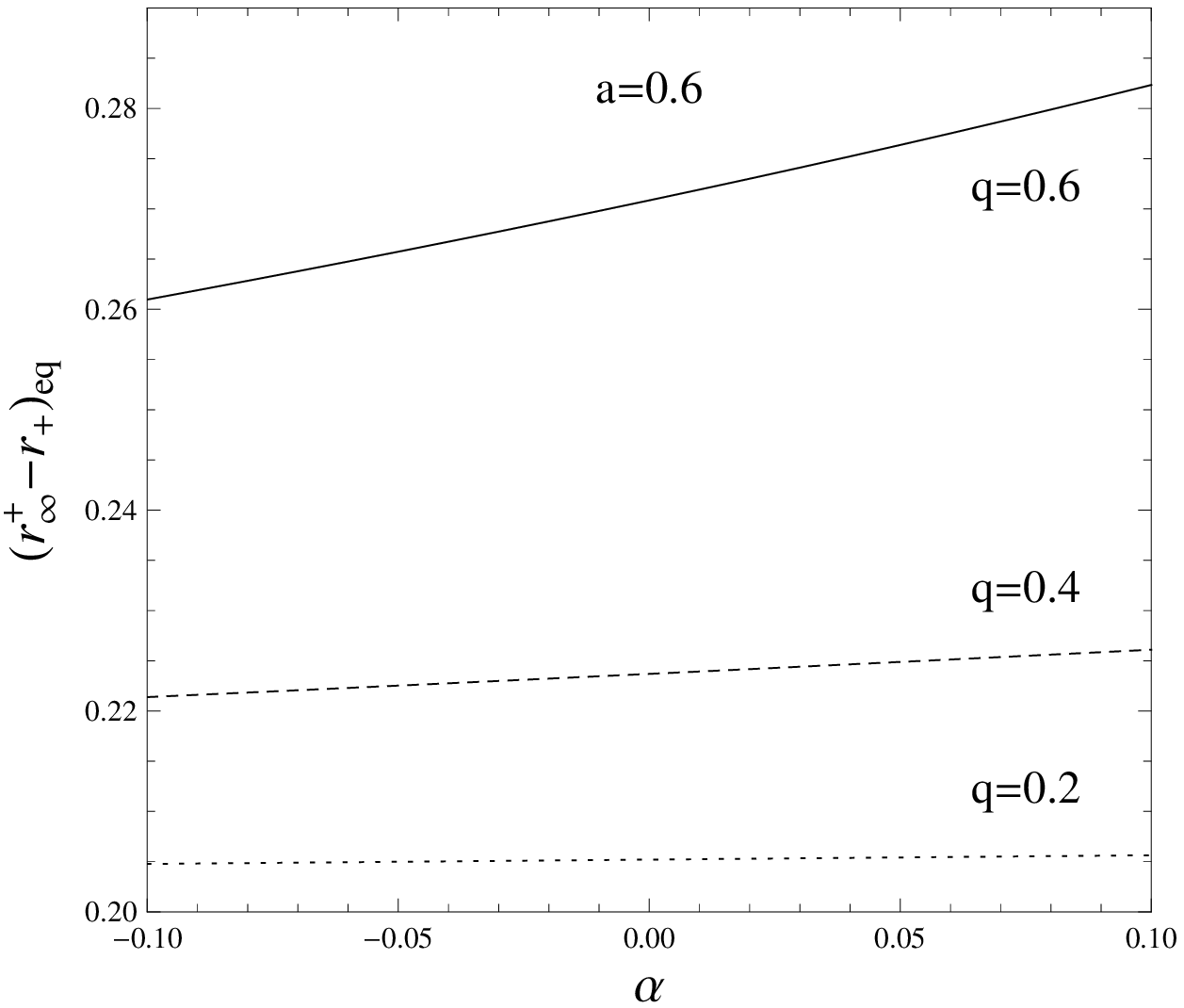}
\caption{The change of the width of ergosphere in the equatorial plane $(r^+_{\infty}-r_+)_{eq}$ with the Weyl coupling parameter $\alpha$ for different $a$ and $q$. Here we set $M=1$.}
\end{center}
\end{figure}
Comparing with the spacetime with the negative Weyl coupling constant (.i.e, $\alpha<0$), the change of the topological properties of the horizons and the infinite redshift surfaces become easier in the
spacetime with the positive Weyl coupling constant (.i.e, $\alpha>0$). For fixed rotation parameter $a$, we also note that with the increasing Weyl corrections, the ergosphere in the equatorial plane becomes thick  for a black hole with $\alpha>0$, but becomes thin in the case with $\alpha<0$, which is shown in figure (9). It means that with the increase of the Weyl corrections the energy extraction become easier in the background of a black hole with $\alpha>0$, but more difficult in the background of a black hole with $\alpha<0$.

\section{summary}

In this paper we present firstly a four-dimensional spherical symmetric black hole with small Weyl corrections and find that the Weyl coupling parameter $\alpha$ affects the radius of the event horizon, Hawking temperature and hawking radiation of the black hole. Moreover, we find that with the increasing Weyl corrections the region of the event horizon existence for the black hole increases for the negative $\alpha$ and decreases for the positive $\alpha$. Moreover, we obtain also a rotating charged black hole with weak Weyl corrections by the method of complex coordinate
transformation. We find that presence of Weyl corrections makes the black hole horizon  as a function of the polar angle coordinate $\theta$, which brings some special properties of the black hole horizon. In particular, the rotating black hole with $\alpha<0$ develops two disconnected topologically spherical
horizons above some critical spin and charge parameters. In the case
with $\alpha>0$, the horizon looks more like a toroidal surface above some critical $a$ and $q$. Comparing with the spacetime with $\alpha<0$, the change of the topological properties of the horizons become easier in the spacetime with $\alpha>0$. We also analyze the dependence of the ergosphere on the Weyl coupling parameter $\alpha$ and find that the ergosphere in the equatorial plane becomes thick for a black hole with $\alpha>0$, but becomes thin in the case with $\alpha<0$. It means that with the increase of the Weyl corrections the energy extraction become easier in the background of a black hole with $\alpha>0$, but more difficult in the background of a black hole with $\alpha<0$.

Finally, we must point out that we here have not discussed the thermodynamic properties of the rotating charged black hole with the Weyl corrections. The main reason is that the radius of black hole horizon is a function of the polar angle coordinate $\theta$, which leads to that the surface gravity $\kappa$ is not a well-defined quantity at the Killing horizon.

\section{\bf Acknowledgments}
This work was  partially supported by the National Natural Science Foundation of
China under Grant No.11275065, the NCET under Grant
No.10-0165, the PCSIRT under Grant No. IRT0964,  the Hunan Provincial Natural Science Foundation of China (11JJ7001) and the construct
program of key disciplines in Hunan Province. J. Jing's work was
partially supported by the National Natural Science Foundation of
China under Grant Nos. 11175065, 10935013; 973 Program Grant No.
2010CB833004.


\vspace*{0.2cm}
\begin{thebibliography}{99}

\baselineskip=0.6 cm \baselineskip=0.6 cm

\bibitem{Born} M. Born and L. Infeld, Proc. R. Soc. A {\bf144}, 425
(1934)
\bibitem{Boillat} G. Boillat, J. Math. Phys. {\bf11}, 941 (1970); {\bf11}, 1482
(1970).
\bibitem{Gibbons}G. W. Gibbons and D. A. Rasheed, Nucl. Phys. B {\bf454}, 185 (1995).

\bibitem{Fradkin} E. Fradkin and A. A. Tseytlin, Phys. Lett. B {\bf163}, 123
(1985); A. Abouelsaood, C. G. Callan Jr., C. R. Nappi,
and S.A. Yost, Nucl. Phys. B {\bf280}, 599 (1987); R. G.
Leigh, Mod. Phys. Lett. A {\bf4}, 2767 (1989); D. Brecher,
Phys. Lett. B {\bf442}, 117 (1998); D. Brecher and M. J.
Perry, Nucl. Phys. B {\bf527}, 121 (1998); A. A. Tseytlin,
Nucl. Phys. B {\bf501}, 41 (1997).


\bibitem{Balakin}  A. B. Balakin and J. P. S. Lemos, Class. Quantum Grav. {\bf22}, 1867 (2005).
\bibitem{Faraoni}  V. Faraoni, E. Gunzig and P. Nardone, Fundamentals of Cosmi
Physis {\bf20}, 121 (1999).
\bibitem{Hehl} F. W. Hehl and Y. N. Obukhov, Lect. Notes Phys. {\bf562}, 479 (2001).


\bibitem{Turner} M. S. Turner and L. M. Widrow , Phys. Rev. D {\bf37} 2743 (1988).
\bibitem{Mazzitelli} F. D. Mazzitelli and F. M. Spedalieri, Phys. Rev. D {\bf52} 6694 (1995).
\bibitem{Lambiase} G. Lambiase and A. R. Prasanna, Phys. Rev. D {\bf70}, 063502 (2004).
\bibitem{Raya} A. Raya, J. E. M. Aguilar and M. Bellini, Phys. Lett. B {\bf638}, 314 (2006).
\bibitem{Campanelli} L. Campanelli, P. Cea, G. L. Fogli and L. Tedesco, Phys. Rev. D {\bf77}, 123002 (2008).

\bibitem{Bamba} K. Bamba and S. D. Odintsov, JCAP {\bf0804}, 024, (2008).
\bibitem{Kim} K. T. Kim, P. P. Kronberg, P. E. Dewdney and T. L. Landecker, Astrophys. J. {\bf355} 29 (1990); K.T. Kim, P. C. Tribble and P. P. Kronberg, Astrophys. J. {\bf379} 80 (1991).
\bibitem{Clarke}  T. E. Clarke, P. P. Kronberg and H. Boehringer, Astrophys. J. {\bf547}, L111 ( 2001).

\bibitem{BIBH1} B. Hoffmann, Phys. Rev. {\bf47}, 877 (1935).
\bibitem{BIBH2} A. Garcia, H. Salazar, and J. F. Plebanski, Nuovo. Cim.
{\bf84}, 65 (1984); M. Demianski, Found. of Phys. {\bf16}, 187
(1986); N. Breton, Phys. Rev. D {bf67}, 124004 (2003).
\bibitem{BIBH3} S. Fernando and D. Krug, Gen. Rel. Grav. {\bf35}, 129 (2003);
T. K. Dey, Phys. Lett. B {\bf595}, 484 (2004); R. G. Cai,
D. W. Pang, and A. Wang, Phys. Rev. D {\bf70}, 124034
(2004); O. Miskovic and R. Olea, Phys. Rev. D {\bf77},
124048 (2008).

\bibitem{Bardeen} J. Bardeen, Proc. of GR5, Tiflis, USSR, 1968
\bibitem{Borde} A. Borde , Phys.Rev. D50 (1994) 3392
 A. Borde , Phys.Rev. D55 (1997) 7615
\bibitem{Beato1} E. Ay\'{o}n-Beato and A. Garc\'{i}a, Phys. Rev. Lett. {\bf80}, 5056 (1998); Phys. Lett B {\bf493}, 149 (2000); Phys. Lett B {\bf464} 25 (1999).
\bibitem{Beato} E. Ay\'{o}n-Beato and A. Garc\'{i}a,  Gen. Rel. Grav. {\bf37}, 635 (2005).


\bibitem{pwbh1} M. Hassaine and C. Martinez,  Phys. Rev. D {\bf75}, 027502 (2007); hep-th/0701058.
\bibitem{pwbh2} H. Maeda, M. Hassaine, C. Martinez, Phy. Rev. D {\bf79}, 044012 (2009).
\bibitem{pwbh3} M. Hassaine and C. Martinez,  Phy. Rev. D {\bf75}, 027502 (2007).
\bibitem{pwbh4} O. Gurtug, S. H. Mazharimousavi, and M. Halilsoy, Phys. Rev. D {\bf85}, 104004 (2012); arxiv: 1010.2340 [gr-qc].


\bibitem{Balakin1}  A. B. Balakin, V. V. Bochkarev and J. P. S. Lemos,  Phys. Rev. D {\bf 77}, 084013 (2008).

\bibitem{Weyl1} A. Ritz and J. Ward,  Phys. Rev. D {\bf79} 066003 (2009).
\bibitem{Drummond}I. T. Drummond and S. J. Hathrell, Phys. Rev. D {\bf22}, 343
(1980).

\bibitem{Dereli1} T. Dereli1 and O. Sert, Eur. Phys. J. C  {\bf71}, 1589 (2011).

\bibitem{Solanki} S. K. Solanki, O. Preuss, M. P. Haugan, A. Gandorfer, H. P. Povel, P. Steiner, K. Stucki, P. N. Bernasconi, and D. Soltau,
Phys. Rev. D {\bf69}, 062001 (2004); O. Preuss, M. P. Haugan, S. K. Solanki, and S. Jordan, Phys. Rev. D {\bf70}, 067101 (2004); Y. Itin and F. W. Hehl, Phys. Rev. D {\bf68}, 127701 (2003).



%\bibitem{Prasanna}A. R. Prasanna, Phys. Lett. A {\bf37}, 337 (1971)
%\bibitem{GW1} G.W. Horndeski, J. Math. Phys. {\bf17}, 1980 (1976)

\bibitem{Wu2011} J. P. Wu, Y. Cao, X. M. Kuang, and W. J. Li, Phys. Lett. B {\bf697}, 153 (2011).
\bibitem{Ma2011} D. Z. Ma, Y. Cao, and J. P. Wu, Phys. Lett. B {\bf704}, 604 (2011).
\bibitem{Momeni} D. Momeni, N. Majd, and R. Myrzakulov, Europhys. Lett. 97, 61001 (2012).
\bibitem{Roychowdhury} D. Roychowdhury, Phys. Rev. D {\bf86}, 106009 (2012); D. Momeni, M. R. Setare, and R. Myrzakulov, Int. J. Mod.
Phys. A {\bf27}, 1250128 (2012); D. Momeni and M. R. Setare, Mod. Phys. Lett. A {\bf26}, 2889 (2011).
\bibitem{zhao2013} Z. X. Zhao, Q. Y. Pan, J. L. Jing,  Phys. Lett. B {\bf719}, 440  (2013).


\bibitem{sb2013} S. Chen and J. Jing, Phys. Rev. D {\bf88}, 064058 (2013).

\bibitem{JNT} E. T. Newman and A. I. Janis, J. Math. Phys. {\bf6} 915 (1965).

\bibitem{JNT1} H. Stephani, \textit{Exact Solutions of Einstein's Field Equations}, (Cambridge University Press, 2003).

\bibitem{BY} J. Brown and J. York, Phys. Rev. D {\bf47}, 1407 (1993).
\bibitem{STZ} S. Bose and T. Z. Naing, Phys. Rev.D {\bf60}, 104027 (1999).
\bibitem{RBM} M. H. Dehghani and R. B. Mann, Phys. Rev.D {\bf64},044003, (2001).
\bibitem{Sheykhi1} A. Sheykhi and M. Allahverdizadeh, Gen. Rel. Grav. {\bf42}, 367 (2010).
\bibitem{Hendi} S. H. Hendi, Prog. Theor. Phys. {\bf124}, 493 (2010).
\bibitem{LBS} L. B. Szabados,
Living Rev. Relativity {\bf 12}, 4 (2009).

\bibitem{TJo} T. Johannsen, D. Psaltis, Phys. Rev. D {\bf83} 124015 (2011).

\bibitem{FLM} F. Caravelli, L. Modesto, Class. Quant. Grav. {\bf27}, 245022
(2010), arXiv:1006.0232 [gr-qc].

\bibitem{CBa} C. Bambi, L. Modesto, Phys. Lett. B {\bf706} 13 (2011).
\end{thebibliography}
\end{document}